\documentclass[a4paper,11pt]{article}
\pdfoutput=1 % if your are submitting a pdflatex (i.e. if you have
\usepackage[letterpaper,top=2cm,bottom=2cm,left=3cm,right=3cm,marginparwidth=1.75cm]{geometry}
\usepackage{amsmath}
\usepackage{graphicx}
\usepackage{bm}
\usepackage{comment}
\usepackage{amssymb}
\usepackage{authblk}

\title{\textbf{Emergent Computations in Trained Artificial Neural Networks and Real Brains}}
\author[,a,b]{N\'estor Parga\footnote{Corresponding author at nestor.parga@uam.es \\\textbf{International Summer School on Intelligent Signal Processing for Frontier Research and Industry} (INFIERI 2021).  Universidad Aut\'onoma de Madrid, Madrid, Spain. 23 August - 4 September 2021.}}
\author[a,b]{Luis Serrano-Fern\'andez}
\author[c]{Joan Falc\'o-Roget}

\affil[a]{Departamento de F\'isica Te\'orica, Universidad Aut\'onoma de Madrid, 28049 Madrid, Spain}
\affil[b]{Centro de Investigaci\'on Avanzada en F\'isica Fundamental, Universidad Aut\'onoma de Madrid, 28049 Madrid, Spain}
\affil[c]{Sano - Centre for Computational Personalised Medicine, 30-054 Krak\'ow, Poland}

\date{}

\providecommand{\keywords}[1]
{
  \small	
  \textbf{\textit{Keywords---}} #1
}

\begin{document}
\maketitle
\flushbottom

\begin{abstract}
Synaptic plasticity allows cortical circuits to learn new tasks and to adapt to changing environments.
How do cortical circuits use plasticity to acquire functions such as decision-making or working memory? Neurons are connected in complex ways, forming recurrent
neural networks, and learning modifies the strength of their connections. Moreover, neurons
communicate emitting brief discrete electric signals. Here we describe how to train recurrent
neural networks in tasks like those used to train animals in neuroscience
laboratories, and how computations emerge in the trained networks. Surprisingly, artificial networks
and real brains can use similar computational strategies.
\end{abstract}

\keywords{Simulation methods and programs, Data processing methods}

%%%%%%%%%%%%%%%%%%%%%%%%%%%%%%%%
%%%%%%%%% INTRODUCTION %%%%%%%%% 
%%%%%%%%%%%%%%%%%%%%%%%%%%%%%%%%
\section{Introduction}
\label{sect:intro}

New computational techniques \cite{sussillo2009generating, laje2013robust, depasquale2016using,depasquale2018full,kim2019simple, miconi2017biologically, song2017reward} enable neural networks to be trained on tasks similar to those used in experiments with behaving animals \cite{romo1999neuronal,padoa2006neurons,kiani2009representation,raposo2012multisensory, jazayeri2015neural,wang2018flexible,de2005neuronal,carnevale2013optimal}.  
Before these techniques became available, a researcher would hypothesize what computations the network should perform to execute the task, and build a network architecture capable of carrying them out. Then, numerical simulations of the model or mean field approximations allowed verifying whether the proposed network model performed the task as desired. This is unsatisfactory, as it does not allow identifying how a neural network could solve these tasks; the models thus constructed only reflect the researcher's intuitions about how the tasks could be performed. In contrast, trained networks provide us with a valuable tool to investigate mechanisms that networks could use to perform the tasks \cite{mante2013context,barak2013fixed, carnevale2015dynamic,song2016training,miconi2017biologically,song2017reward, chaisangmongkon2017computing,wang2018flexible,yang2019task,kim2021strong, serrano2022TID}. This approach enables to formulate hypotheses, draw predictions and devise analysis methods that could help to guide experiments.

The aim of this paper is to discuss ways to build neural network models to investigate how the brain could solve cognitive tasks. Accordingly, the focus will be on the biological plausibility of the models and techniques used to formulate them. In this regard, it is necessary to distinguish the roles of plausibility in each of the two stages of the program: (1) the biological realism of the learning paradigm and the learning rules used for training and (2) the biological realism of the trained networks. We emphasize here that, once the network has been optimized, the training is interrupted and simulated data are collected to analyze how the network solves the task. So, even if the network has been trained with update rules that are difficult to justify from a biological perspective, the trained network can be used to investigate how a certain type of network model can solve a given cognitive task. 
This observation is relevant because there are currently no methods to train neural networks with biologically realistic architectures, using learning schemes that are also biologically plausible. As we will see below, from a biological standpoint, it would be desirable to have the ability to build models of recurrent networks of spiking neurons, trained with reward-based algorithms and Hebbian-type rules \footnote{Note that here we will not deal with the training of neural networks to predict neural activity patterns, as in, e.g. \cite{yamins2016using}}. However, all current techniques employed to model cognitive tasks, only consider some of these basic requirements.

This work is organized as follows: after introducing several basic concepts (Sect. \ref{sect:basic-material}) we present the most widely used learning paradigms, both in Machine Learning (ML)  and in Neuroscience: supervised learning (Sect. \ref{sect:supervised}) and reinforcement learning (Sect. \ref{sect:RL}). In both cases we discuss techniques that have proven to be effective in training neural networks and that can be taken as a basis for future advances and the development of novel methods. In Sect. \ref{sect:biological} we go on to discuss the achievements and weaknesses of current methods for constructing neural network models that are acceptable from a biological perspective.

%%%%%%%%%%%%%%%%%%%%%%%%%%%%%%%%
%%%%%%%% BASIC MATERIAL %%%%%%%% 
%%%%%%%%%%%%%%%%%%%%%%%%%%%%%%%%
\section{Basic Material}
\label{sect:basic-material}

We start by introducing basic concepts of several neuroscience and ML topics, necessary to make the content of this paper more accessible. After giving the fundamental notion of task, we describe the elements required to build a neural network model, i.e., model neurons and network architectures. We then go through several issues related to the training of these networks: learning paradigms, the credit assignment problem, Hebbian-like learning and update rules.

\subsection{Tasks}
\label{subsect:Tasks}

\begin{figure}[h]
    \centering
    \includegraphics[scale=0.35]{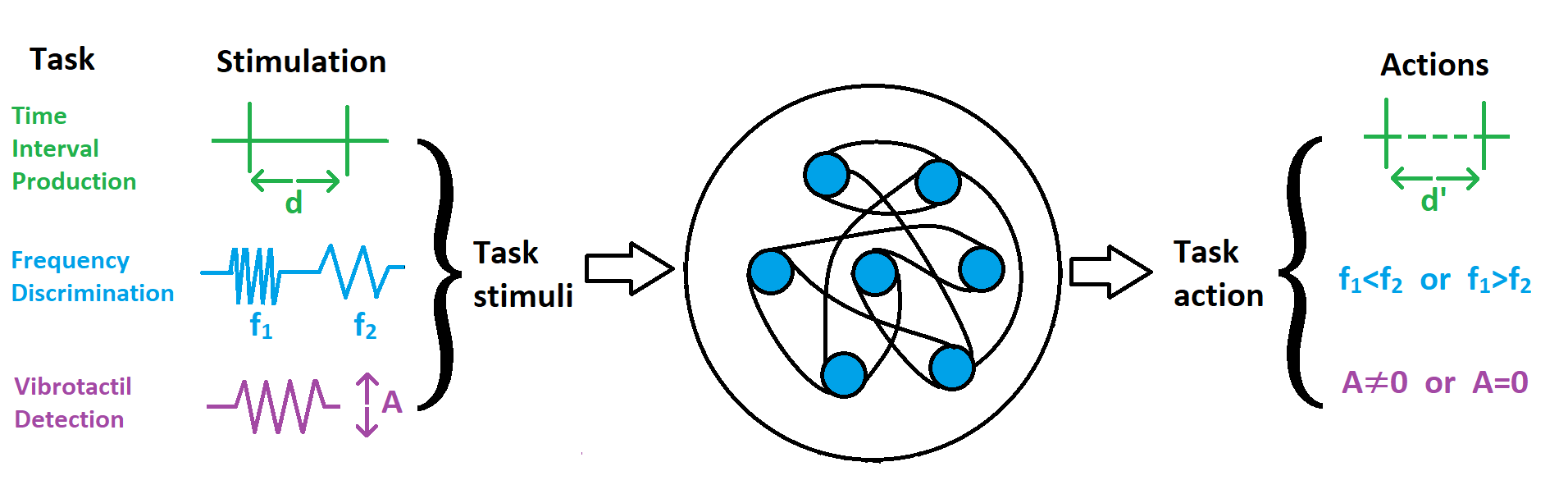}
    \caption{{\bf A task as a mapping from stimuli to actions}. A neural network performing a task associates stimuli to actions. Three tasks are shown: Time interval reproduction task (green), in which the network receives a time interval, $d$, and reproduces it, $d^\prime$. Frequency discrimination task (blue) where the network receives two stimuli separated by a delay interval and has to report which is higher. A detection task (purple) in which the network receives a stimulus of variable amplitude $A$ and reports whether it is zero or not.}
    \label{Fig:task}
\end{figure}

\noindent
A task is a mapping from stimuli to actions. Figure \ref{Fig:task} illustrates this notion with three mappings, corresponding to tasks that can be investigated with electrophysiology experiments: frequency discrimination task \cite{romo1999neuronal}, time interval reproduction task \cite{jazayeri2015neural} and a vibrotactile detection task \cite{de2005neuronal,carnevale2013optimal}.

\noindent As Figure \ref{Fig:task} indicates, here we will focus on how mappings can be implemented by neural networks, which receive inputs (stimuli) and generate outputs (actions). The network's responses depend on parameters (e.g., the synaptic couplings or weights), some of which will be assumed to be plastic, i.e. their values can be optimized in such a way that the input-output relationship produced by the network corresponds to that dictated by the task under consideration. The learning paradigm is the theoretical framework used to train the network to perform a task.  The learning rule (or update rule) is the precise prescription by which the network parameters are modified during learning. As we describe next, the model neurons (or units) that compose the network can be taken in a variety of ways, from rather artificial models to more biologically realistic ones. Regarding the network architecture, we will consider two basic types, feedforward (Figure \ref{Fig:Feedfordwar_&_Recurrent}, left) and recurrent networks (Figure \ref{Fig:Feedfordwar_&_Recurrent}, right).

\subsection{Model neurons}
\label{subsect:Model_Neurona}

Rate neurons are given by continuous variables; the activity input to unit $i$, $x_i(t)$, is 

\begin{equation}
	\tau_x\frac{dx_i(t)}{dt} = - x_i(t) + I_{ext},
 \label{eq:dyn_rate}
\end{equation}

\noindent where $I_{ext}$ is an external current and $\tau_x$ is the characteristic time \cite{dayan2005theoretical}. However, neurons emit spikes, which are discrete events localized in time. Even so, biological studies very often describe neurons with continuous variables; in these cases the firing rate of the neuron (number of spikes emitted by unit time) is a convenient nonlinear function of the input activity, $r(x_i)$. This is usually taken as a threshold linear function, $r(x_i)= [x_i]_+ = max(0; x)$, or a sigmoid function, or (for analytical convenience) even as $r(x_i)= tanh(x_i)$, even when $r(x_i)$ should be a positive quantity. 

A more realistic neuron model, emitting spikes, is the Leaky-Integrate-and-Fire (LIF) model. This model describes the membrane potential $V_i(t)$ of neuron $i$ as

\begin{equation}
	\tau_m\frac{dV_i(t)}{dt} = V_{rest} - V_i(t) + I_{ext},
 \label{eq:dyn_LIF}
\end{equation}

\noindent where $\tau_m$ is the characteristic membrane time constant and $I_{ext}$ is an external current to the neuron \cite{dayan2005theoretical}. A LIF neuron produces a spike when its membrane potential reaches a threshold value, $V_{th}$. Then, the potential is reset to $V_{rest} < V_{th}$. Once reset, the neuron cannot produce a new spike during a refractory time interval, $t_{ref}$.

\subsection{Model networks}
\label{subsect:Model_Neworks}

\subsubsection{Feedforward Neural Networks}
\label{subsubsect:Model_Neworks_FFNNs}

\begin{figure}[h]
    \centering
    \includegraphics[scale=0.30]{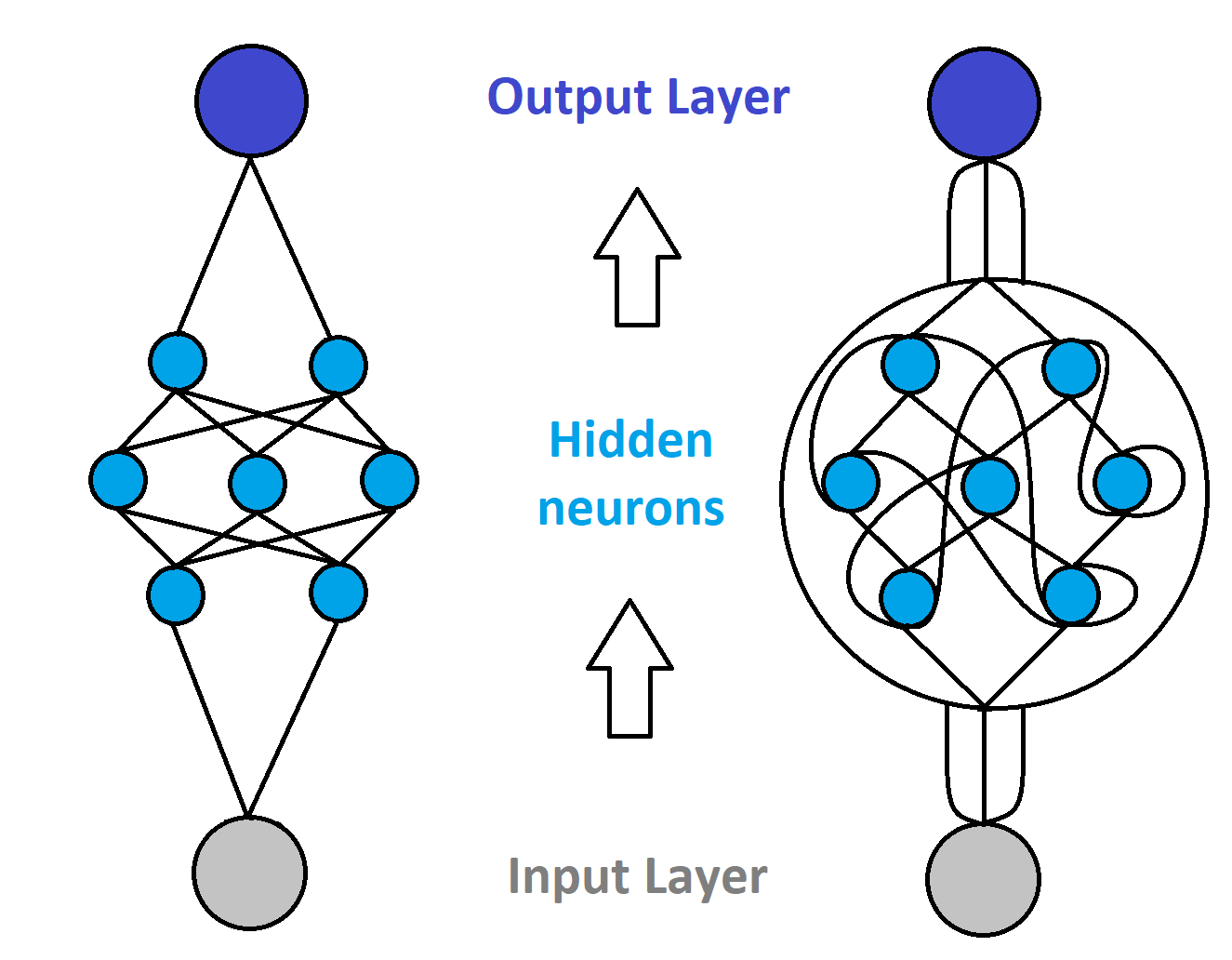}
    \caption{{\bf Feedforward and Recurrent Network Architectures}.  {\bf Left}. Feedforward neural network with an input layer (big grey circle), followed by layers of hidden neurons (small blue circles) and an output layer (big dark blue circle). Layers are connected only forward. {\bf Right}.  Recurrent neural network architecture.}
    \label{Fig:Feedfordwar_&_Recurrent}
\end{figure}

In feedforward neural networks (FFNNs) the units are organized according to a sequence of layers (Figure \ref{Fig:Feedfordwar_&_Recurrent}, left). An input {\boldmath{$F^{(in)}$}} applied on the input layer, flows through the $L$ hidden layers via a set of weights {\boldmath{$W^{l}$}} ($l=0,\ldots,L$), reaching the output layer (Figure \ref{Fig:connectivity_rate_&_Spike_RNNs}, left). The weight $W^{l-1}_{i,j}$ connects unit $j$ on layer 
$l-1$ with unit $i$ on layer $l$. The activity $x^{l}_{i}$ of the latter is 

\begin{equation} \label{eq:forward_activity}
    x^l_i = r \big( \sum_{j=1}^{N_l} W^{l-1}_{i,j} x^{l-1}_j + b^{l-1}_i \big)
\end{equation}

\noindent  where $N_l$ is the number of neurons on the $l$-th layer and $r(\cdot)$ is a nonlinear function (e.g., a sigmoid function). Importantly, the activity of the $i$-th input unit is $x^0_i \doteq F_i^{(in)}$. The biases $b^{l}_i$ are also a set of modifiable inputs that might be interpreted as external currents, $I^l_{ext,i}$ (Eqs. \ref{eq:dyn_rate} and \ref{eq:dyn_LIF}).  Recall that the activity of each neuron on the $l$-th layer is dependent on the activity of all the neurons in the previous layers. By iterating through all the $L$ hidden layers, the input {\boldmath{$F^{(in)}$}} is forwarded through the network being finally mapped to the output $z$ (for simplicity, in the equations we assume a single unit in the output layer)

\begin{equation}
 \label{eq:output_FFNN}
    z \doteq r \big( \sum_{j=1}^{N_L} W^{L}_j x^{L}_j + b^{L} \big).
\end{equation}

\noindent Note that $z$ is a highly nonlinear function of couplings, biases and inputs, $z=f\big(${\boldmath{$W^{0}$}},\ldots,{\boldmath{$W^{L}$}}; {\boldmath{$b^{0}$}},\ldots,{\boldmath{$b^{L}$}};  {\boldmath{$F^{(in)}$}}$\big)$. When the number of layers is large, we speak of deep networks \cite{lecun2015deep}.

\subsubsection{Recurrent Neural Networks of Rate Neurons}
\label{subsubsect:RNNs_rate}

The brain solves cognitive tasks (Sect. \ref{subsect:Tasks}) in areas that can be described as networks with recurrent connectivity, in which neurons can connect to each other in any direction  (Figure \ref{Fig:Feedfordwar_&_Recurrent}, right). Being this architecture biologically more realistic, we will focus here on recurrent neural networks (RNNs), distinguishing between RNNs of continuous-variable units (rate neurons) and RNNs of spiking neurons (LIF). We start with the former; in a RNN with $N_{in}$ input neurons and $N$ hidden (internal) neurons (Figure \ref{Fig:connectivity_rate_&_Spike_RNNs}, middle), each hidden unit $i$ is described by an activation variable $x_i$ evolving as

\begin{equation} 
	\tau_x\frac{dx_i(t)}{dt} = - x_i(t) +  g \sum_{j=1}^{N} W_{i,j} r(x_j(t)) + \sum_{l=1}^{N_{in}} W_{i,l}^{(in)} F_l^{(in)}(t),  \label{eq:dyn_rate_RNN}
\end{equation}

\noindent
where the $N \times N$ connection matrix {\boldmath{$W$}} is often built randomly from a Gaussian distribution of zero mean and variance $1/N$. $r(x_i)$ is the rate activity of unit $i$, usually taken as $r(x_i)= tanh(x_i)$ or as a threshold linear function, $r(x_i)= [x_i]_+ = max(0; x)$. Note that this model assumes that there is a single type of recurrent connections. $W_{i,l}^{(in)}$ denotes the synapse that introduces the input signal $F_l^{(in)}$ (from input unit $l$ to unit $i$) into the network. It has dimensions $N \times N_{in}$  and it is built, e.g., from a uniform distribution. $\tau_x$ is the relaxation time constant of neuron activation. The output neuron has activity  $z(t)=\sum_{j=1}^{N} W_{j}^{(out)} r(x_j(t))$ (we will always assume a single output or readout neuron, although the generalization to more neurons is trivial). The constant $g$ is a gain factor that can be used to control the dynamic regime of the network model before training. When $g>1$ the activity of the  firing-rate model network in Eq. \ref{eq:dyn_rate_RNN} exhibits chaotic dynamics \cite{sompolinsky1988chaos}.

\subsubsection{Recurrent Neural Networks of Spiking Neurons}
\label{subsubsect:RNNs_spiking}

\begin{figure}[h]
    \centering
    \includegraphics[scale=0.3]{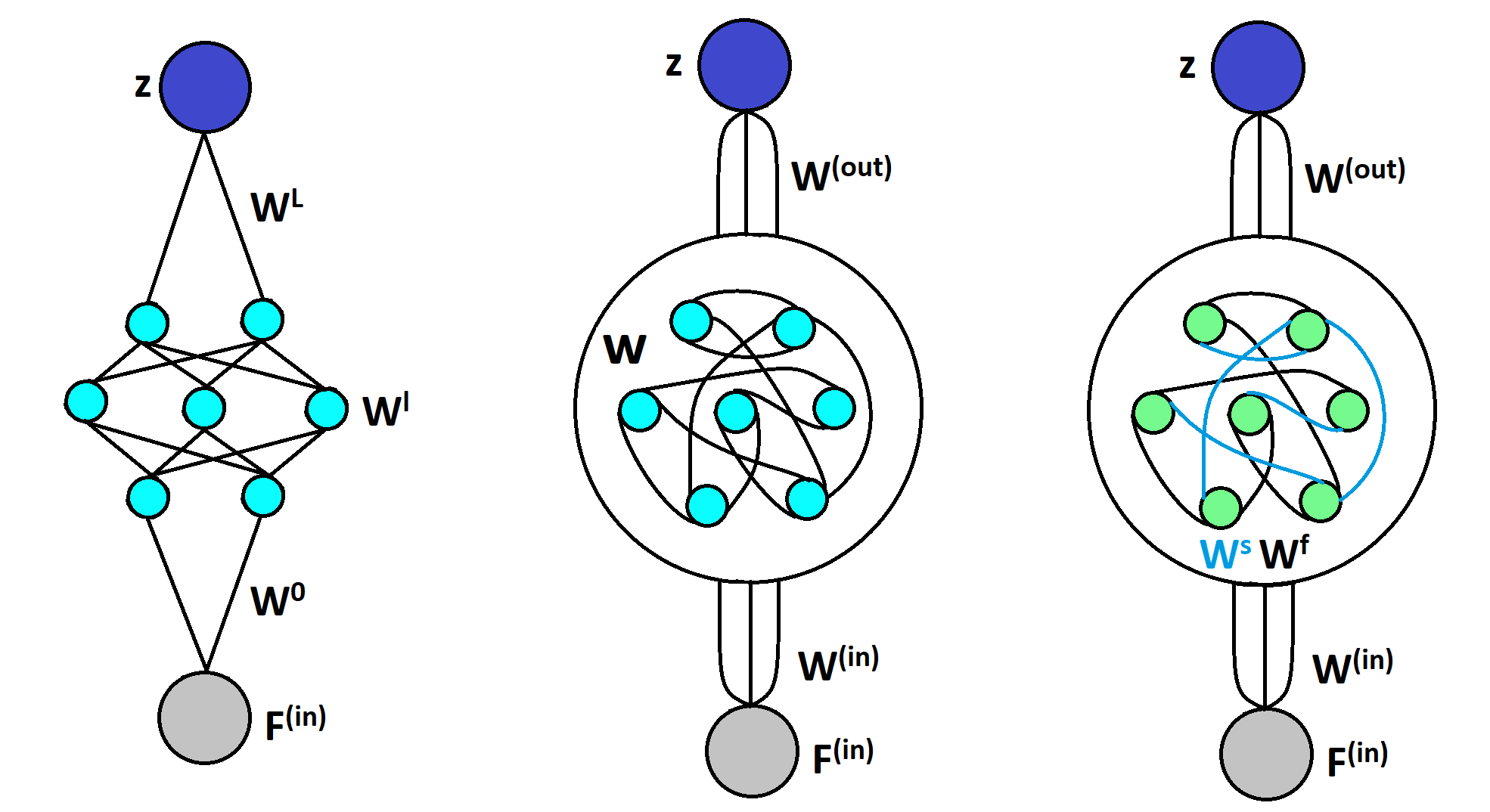}
    \caption{{\bf Connectivity in Feedforward Neural Networks, rate RNNs and spiking RNNs}. {\bf Left}: A Feedforward Network of continuous variable units, which receive an input signal, {\boldmath{$F^{(in)}$}}, from a layer of input units (grey circle) and generated an output current, $z$, read by an output layer (dark blue circle). The $L$ hidden rate layers (cyan circles) are connected to each other through forward synapses, $W^l$ ($l=1,...,1-L$) (black lines), and to both input and output layers through $W^{0}$ and $W^L$, respectively. {\bf Middle}: A rate RNN with recurrent couplings $W$, (black lines) which receive its input signal through the input connections {\boldmath{$W^{(in)}$}}. The output current, $z$, is read by the output layer (which is connected to the hidden layers through the synapses $W^{(out)}$). {\bf Right}: A RNN LIF neurons. The hidden LIF neurons (green circles) are connected to each other through two types of recurrent synapses, the fast ({\boldmath{$W^f$}}, black lines) and the slow ({\boldmath{$W^s$}}, blue lines) synapses. They convey fast ({\boldmath{$f$}}$(t)$) and slow ({\boldmath{$s$}}$(t)$) currents, respectively.}
    \label{Fig:connectivity_rate_&_Spike_RNNs}
\end{figure}

\noindent
We considered spiking recurrent neural networks (sRNNs) of leaky integrate-and-fire (LIF) neurons (Figure \ref{Fig:connectivity_rate_&_Spike_RNNs}, right). The dynamics of the membrane potential of neuron $i$, $V_i(t)$, is obtained from Eq. \ref{eq:dyn_LIF} by adding the contributions from the recurrent currents 

\begin{equation}
	\tau_m\frac{dV_i(t)}{dt}=V_{rest} - V_i(t) + g \sum_{j=1}^{N} [W_{i,j}^s s_j(t) + W_{i,j}^f f_j(t) ] + \sum_{l=1}^{N_{in}} W_{i,l}^{(in)} F_l^{(in)}(t) + I_{ext},
 \label{eq:dyn_LIF_RNN}
\end{equation}

\noindent
$\tau_m$  is the membrane time constant, $V_{rest}$ is the rest potential and $g$ is a gain factor. {\boldmath{$W^{(in)}$}} are the input weights (defined, e.g., from an uniform distribution) through which the input signal, {\boldmath{$F^{(in)}$}}, is introduced into the network. The matrices {\boldmath{$W^s$}} and {\boldmath{$W^f$}} are hidden recurrent connections associated with slow ({\boldmath{$s$}}$(t)$) and fast ({\boldmath{$f$}}$(t)$) synaptic currents, respectively. The dynamics of these currents are

\begin{equation}
	\tau_s\frac{ds_j}{dt}=-s_j(t) \hspace{5mm}	\textrm{and} \hspace{5mm} \tau_f\frac{df_j}{dt}=-f_j(t),
 \label{eq:dyn_syn_curr}
\end{equation} 

\noindent
As their names indicate, their decay times are long (e.g., $\tau_s = 100$ ms) or short (e.g., $\tau_f = 5$ ms).    

Once a neural network has been built, the next step is to select a learning paradigm and learning rule to train the network to perform a task.

\subsection{Learning Paradigms, Plasticity and Update of Synaptic Connections}
\label{subsect:plasticity}

\subsubsection{Learning} 
\label{subsubsect:learning}

Intuitively, learning can be understood as the constant promotion of changes in the structure of a system so that it is able to aggregate similar concepts and to differentiate dissimilar ones \cite{lynn2020humans}. The concept notion is rather broad, but in the context of the present work, it can be simplified to the inputs that a network processes in order to output the desired behavior\footnote{For example, correctly classifying two different inputs to the same class.}. Formally, learning implies the optimization of some weight-dependent function that quantifies the performance of the system, so that when the performance is best, this function lies at a maximum (or a minimum) \cite{bishop2006pattern}. A \textit{learning rule} proposes changes in the weights by means of an unambiguous mathematical statement that can be repeated as many times as required without the need of continuous supervision.

\subsubsection{Learning Paradigms}
\label{subsubsect:Learning_paradigms}

In ML, the goal is to develop rules that autonomously provide a general mapping from inputs to outputs. However, the methods to \textit{learn} the input-output maps depend on the nature of the \textit{signal} or \textit{feedback} available to the system \cite{bishop2006pattern}. Traditionally, these approaches are grouped into three categories:

\begin{itemize}
    \item \textit{Supervised learning} applies to cases where the true output for a given input is accessible during the learning phase. In regression tasks, inputs and outputs exist in a continuous domain and the goal is to predict a sufficiently accurate estimate of their unknown relationship. In classification tasks, several inputs are mapped to the same output or label.     
    
    \item \textit{Unsupervised learning} broadly refers to inferring underlying structure in the input data without any reference to specific training signals. It can be a goal in itself (i.e., discovering hidden patterns in data) or a means towards an end (i.e., feature learning). One possible theoretical framework for unsupervised learning consists in the maximization of information that the outputs have on the inputs \cite{linsker1988self,atick1990towards,nadal1994nonlinear}. %\cite{linsker1988self,atick1990towards,plumbley1993efficient,nadal1994nonlinear}.    
    Another way to proceed is to minimize the redundancy in the output \cite{barlow1961coding}. However, it was shown that, under some conditions, these two methods are equivalent, that is, the independent components can be obtained by maximizing the mutual information \cite{nadal1994nonlinear}. It should be noted that there is a duality relationship \cite{nadal1994duality} between machines that adapt their weights to maximize information transmission (unsupervised learning) and machines that adapt their weights to maximize information storage (supervised learning).
    
    \item \textit{Reinforcement learning} lies somewhere in the middle. Although the targets are not unambiguously known during the training phase, they can be discovered through a process of trial and error. During learning, an agent interacts with a dynamic environment by visiting states, taking actions 
    %(e.g., selecting a movement in a particular chess position) 
    and receiving rewards. The goal is to maximize this reward signal which, in turn, provides the feedback (i.e., teaching signal) that permits the learning of the task at hand.
\end{itemize}

This paper is mainly devoted to supervised (Sect. \ref{sect:supervised}) and reinforcement learning (Sect. \ref{sect:RL}).

\subsubsection{The Credit Assignment Problem}
\label{subsubsect:CAP}

Determining which parameters (synaptic weights) should be credited for a good or bad outcome is a difficult issue called the {\em credit assignment problem} (CAP) \cite{richards2019dendritic,minsky1961steps}. Neuroscience and ML attempt to solve the CAP from different perspectives. While ML seeks to achieve a system that performs the desired task efficiently (Sect. \ref{subsubsect:Update_rules_AI}), neuroscience looks for solutions that could admit biological implementation (Sect. \ref{subsubsect:Hebb}). A basic issue is the choice of learning rules. We first refer to what kind of rules neuroscience is looking for.

\subsubsection{Hebbian Learning Rule} 
\label{subsubsect:Hebb}

According to Donald O. Hebb, learning produces changes in the strength of the synaptic connections between neurons. In 1949 he formulated a basic notion of how connections are modified by learning \cite{hebb2005organization}.
 Basically, Hebb postulated that if cell 1 repeatedly takes part in firing cell 2, then the strength of the connection from neuron 1 to 2
should increase.
A way to express this notion mathematically is that a synaptic coupling ${\tilde W_{i,j}}$ undergoing Hebbian modifications changes as

\begin{equation} \label{eq_Hebbian_Rule}
\Delta {\tilde W_{i,j}} = \eta  X_{i} Y_{j},
\end{equation}

\noindent where $\eta$ is the learning rate and $Y_{j}$ and $X_{i}$ are variables that describe the activity of the pre- and post-synaptic neurons, respectively. Throughout this work a tilde indicates a plastic synapse, generically, regardless of whether the modifications are Hebbian or not. Note that strong simultaneous activity in the two neurons could greatly strengthen the synaptic efficacy. The persistence of concurrent activities of the two neurons when the network has responded correctly, can be interpreted as an assignment of credit to the synapse for that response. In practice, determining which synapses should be credited for correct or wrong responses is not trivial. In general terms, when searching for solutions to the CAP in biologically-inspired neural networks, synaptic connections are modified  with Hebbian-like rules, or generalizations that introduce features such as modulations by reward or context  \cite{roelfsema2018control,richards2019dendritic} (Sects. \ref{Miconi_2017} and \ref{sect:biological}). For a recent review on synaptic plasticity see, eg., \cite{magee2020synaptic}.

\subsubsection{Learning Rules in Machine Learning}
\label{subsubsect:Update_rules_AI} 

In ML there is more freedom in the choice of the learning (or update) rule. In supervised learning, the most popular rules are backpropagation (for networks with feedforward architecture) \cite{rumelhart1986learning} and backpropagation-through-time \cite{werbos1988generalization}, a version of this algorithm suitable for RNNs,  (see Sect. \ref{subsect:supervised}). In reinforcement learning, the REINFORCE algorithm \cite{williams1992simple} is a prominent one (see Sect. \ref{subsect:RNN_implementation_RL}). Despite their apparent lack of biological plausibility, given the success of these algorithms even in neuroscience problems, there is considerable activity to find approximations that could be implemented biologically (Sect. \ref{sect:biological}). 

%%%%%%%%%%%%%%%%%%%%%%%%%%%%%%%
%%%%% SUPERVISED LEARNING %%%%% 
%%%%%%%%%%%%%%%%%%%%%%%%%%%%%%%
\section{Supervised Learning} \label{sect:supervised}

In supervised learning, any function $C_{\tilde{W}}$  of the distance between the target $z_{tgt}$ and the network output $z$ is, in principle, valid to optimize the plastic weights {\boldmath{$\tilde{W}$}}. A common choice for such a \textit{cost function} is the mean squared distance between the target and the output.

\subsection{Backpropagation} 
\label{subsect:supervised}

For the FFNN all the weights $W^l_{i,j}$ are plastic and are chosen to optimize the cost function

\begin{equation} \label{eq:cost_backprop}
C^{BP}_{W^{bp}} = \frac{1}{N_s} \sum_{i=1}^{N_s} \big(z_i-z_{tgt,i} \big)^2,
\end{equation}

\noindent  where ${\boldmath{W^{bp}}}$ denotes the set of all the weights and the sum is over the $N_s$ associations present in the data. Rumelhart et al. proposed to train these networks by modifying the weights according to the steepest direction of change of this cost function  \cite{rumelhart1986learning}. Using the chain rule to compute the gradient of $C^{BP}_{W^{bp}}$ for a multilayer architecture (Figure \ref{Fig:connectivity_rate_&_Spike_RNNs}, left) one obtains the backpropagation update rule

\begin{equation} \label{eq:backprop}
\Delta W^l_{i,j} \propto \frac{\partial C^{BP}_{W^{bp}}}{\partial W^l_{i,j}}  = \frac{\partial C^{BP}_{W^{bp}}}{\partial z} \frac{\text{d} z}{\text{d} W^l_{i,j}},
\end{equation}

\noindent ($l=0,...,L$), where the second factor is the total derivative of the network's output $z$ (Eq. \ref{eq:output_FFNN}) \textit{w.r.t} the desired weight. For simplicity, the biases $b^l_{i}$ in Eq. \ref{eq:forward_activity} are incorporated in {\boldmath{$W^{bp}$}} and updated accordingly \cite{bishop2006pattern}. Note that non-zero values of the cost function indicate an error that is backpropagated from the output unit through all layers to the first, assigning a credit to each synaptic weight, that is, this algorithm implements  \textit{error backpropagation}. 
%%% BPTT
These authors also acknowledged the possibility to train RNNs, but it was Werbos who formalised the idea \cite{werbos1988generalization}. For that, time has to be unfolded as a multilayer feedforward neural network on which errors can be backpropagated. Backpropagation succesfully and efficiently solves the CAP for artificial feedforward networks. Although there are several reasons to think that it is hard to implement in neural circuits \cite{lillicrap2019backpropagation,lillicrap2020backpropagation,whittington2019theories}, there have been recent efforts to approximate it by biologically plausible means (Sect. \ref{sect:biological}). 

While here we will focus on how to train neural networks, a related issue that has been studied over the years is the generalization capability of layered networks (See e.g., \cite{mato1992generalization,wu2017towards}).

\subsection{Reservoir Learning}
\label{subsect:Reservoir_Learning}

\noindent  A simple way to get an RNN to solve a task is to train only the weights {\boldmath{$ W^{(out)}$}} that converge to the output neuron. The cost function to be minimized with respect to those weights is 

\begin{equation} \label{eq:Reser-Learn}
C_{ W^{(out)}}^{RL} = \langle (z(t)-z_{tgt}(t))^2
\rangle.
\end{equation}

\noindent  The recurrent network acts as a reservoir that is perturbed by the input, which is why this type of training methods are known as reservoir learning 
\cite{maass2002real,jaeger2002adaptive}. 

This method achieves good results, but it has some limitations. Jaeger and Haas used the target not only as a teacher that cues the correct response to the output neuron, but also injected it into the recurrent network itself \cite{jaeger2004harnessing}. Although this proposal is interesting, from the standpoint of training models of cortical networks it presents some problems. First, it is not biologically realistic; a more convincing alternative to feedback the target signal would be to inject the activity of the output neuron, which, as the network gradually learns the task, should approximate the target increasingly better. In addition, feeding back the target prevents the production of fluctuations about this activity pattern, which could be useful for training \cite{sussillo2009generating}. Another limitation arises because training does not begin from a regimen of irregular or chaotic activity. Instead, as was discussed above, models of spontaneously active neural circuits typically exhibit chaotic dynamics, or are in an irregular regime \cite{van1996chaos}. A chaotic activity prior to training is crucial for trial-to-trial variability. Without chaotic dynamic the network could overfit its parameters during the training. In other words, the network would be memorizing the answer of the task instead of understanding why the reason for said response.

\subsubsection{FORCE Algorithm}
\label{subsubsect:FORCE}

From the above discussion we conclude that: 1)  it would be useful for the network to receive information about the activity of the output neuron; 2) it would be convenient to start training from an initial condition with the network activity in a chaotic regime, or at least sufficiently irregular so that the network is able to widely explore the space of possible states. Sussillo and Abbott \cite{sussillo2009generating} found a method to train networks that meets these two requirements: the FORCE algorithm. First, they noticed that to obtain a small error from the first training trial learning should be fast. Then, they considered that a signal of sufficient amplitude and frequency induces a transition to from a chaotic to a non-chaotic state \cite{molgedey1992suppressing,bertschinger2004real,rajan2010stimulus}. Since the network is being driven through the feedback pathway by a signal approximately equal to the target function, this signal can produce the desired transition. 
In this way learning can take place in the absence of chaotic activity, even though prior to learning the network is chaotic and afterwards there may exist additional chaotic trajectories.

\begin{figure}[h]
    \centering
    \includegraphics[scale=0.3]{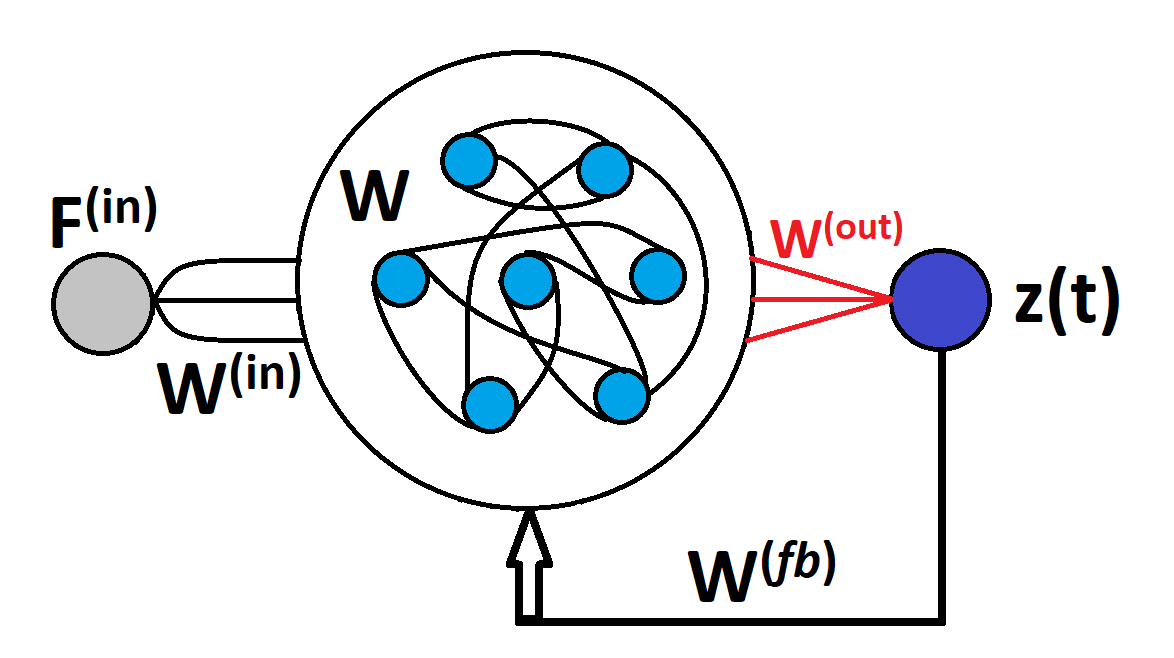}
    \caption{{\bf FORCE scheme}. Feedback to a Recurrent Neural Network (RNN) from the readout neuron layer (dark blue circle) and through the synapses {\boldmath{$W^{(fb)}$}}. The output signal, $z(t)$, returned inside of the RNN, becomes approximately equal to the target function. This signal is received by $N$ hidden neurons (blue circles) connected to each other by non-plastic (black lines) weights, {\boldmath{$W$}}. Furthermore, an input signal, {\boldmath{$F^{(in)}$}}, is introduced to the RNN through the couplings {\boldmath{$W^{(in)}$}}. Finally, the RNN sends its internal current to the readout layer through the plastic (red lines) output weights, {\boldmath{$W^{(out)}$}}, thus completing the circuit. }
    \label{Fig:ReservoirLearning}
\end{figure}

\noindent In short, FORCE  (First-Order Reduced and Controlled Error) works by quickly reducing the error and keeping it low. In others words, FORCE first looks for quickly stabilizing the rates and subsequent weight changes are devoted to stabilizing the weight matrix. To see how to implement this idea, let us go back to the equation for the dynamics of rate RNNs (Eq. \ref{eq:dyn_rate_RNN}) and inject the network output $z(t)$ as a feedback current to the hidden neurons (Figure \ref{Fig:ReservoirLearning}),

\begin{equation}
	\tau_x\frac{dx_i(t)}{dt} = - x_i(t) +  g \sum_{j=1}^{N} W_{i,j} r(x_j(t)) + \sum_{l=1}^{N_{in}} W_{i,l}^{(in)} F_l^{(in)}(t) + W_i^{(fb)} z_{}(t).  \label{eq:dyn_rate_RNN_fback}
\end{equation}

\noindent Note that the feedback term can be written as $\sum_{j=1}^{N} W_i^{(fb)} W_{j}^{(out)} r_j^{}$, which can be interpreted as a  change of the recurrent couplings {\boldmath{$W$}} given by a rank-one matrix. During training, the plastic weights are updated with the Recursive Least Squares algorithm (see details in Appendix \ref{app:RLS}). The recursive nature of this algorithm allows to quickly and effectively find a low error between the network output signal and the desired target signal, guaranteeing activity stability.

\subsection{Targets for Training the Recurrent Connections}

\subsubsection{Laje and Buonomano Algorithm}
\label{subsubsect:Laje_Buonomano}

Although successful in training rate RNNs, the FORCE algorithm cannot adapt the recurrent connections. This is a disadvantage, as fully adaptive RNNs could have a higher learning capacity. 
Building on two previous developments in training rate RNNs \cite{sussillo2009generating,jaeger2004harnessing}, Laje and Buonomano \cite{laje2013robust} developed a novel training algorithm that optimizes not only the readout weights but also the recurrent hidden connections. The key idea was to define targets for the activities of the hidden neurons. 

To do this, one premise was that the rich dynamics of random chaotic networks \cite{sompolinsky1988chaos} can provide a basis for training the network on many tasks. However, the sensitivity to noise and initial conditions of chaotic rate RNNs would make it difficult to use them for training purposes. Laje and Buonomano found an ingenious way out for this problem and for obtaining internal targets for the recurrent connections. In short, to train the recurrent weights in the model network defined by Eq. \ref{eq:dyn_rate_RNN} (they added an independent noise term on the right), they considered an auxiliary rate RNN (the target-generating network) and  arbitrarily selected a typical time-varying activity pattern  ($R_i(t)$, $i=1,\ldots,N$) produced by this network. Then, they used $R_i(t)$ as targets for the activities $r_i(t)$ of the hidden units in the original RNN (the task-performing network). That is, they found the recurrent weights  {\boldmath{$W$}} that minimized 

\begin{equation} \label{eq_cost_LB}
C_{W}^{LB} = \langle \sum_{i=1}^N \big( r_i(t) - R_i(t) \big)^2 \rangle,
\end{equation}

\noindent  where the expectation is over all the times in the trial. The target-generating network operated in the chaotic regime, it was completely deterministic (Eq. \ref{eq:dyn_rate_RNN}, no noise was injected in this network) and the recurrent connections were both random and fixed. The training rule was Recursive Least Squares (Appendix \ref{app:RLS}). After learning, activities from the task-performing network matched those of a chaotic target-generating network but, interestingly, exhibiting a non-chaotic and more stable behavior. Once the task-performing network was stabilized, its readout weights  {\boldmath{$ W^{(out)}$}} were adjusted to minimize the cost function in Eq. \ref{eq:Reser-Learn}.

\subsubsection{full-FORCE Algorithm in Rate RNNs}
\label{subsubsect:targets_full-FORCE}
 
\begin{figure}[h]
    \centering
    \includegraphics[scale=0.3]{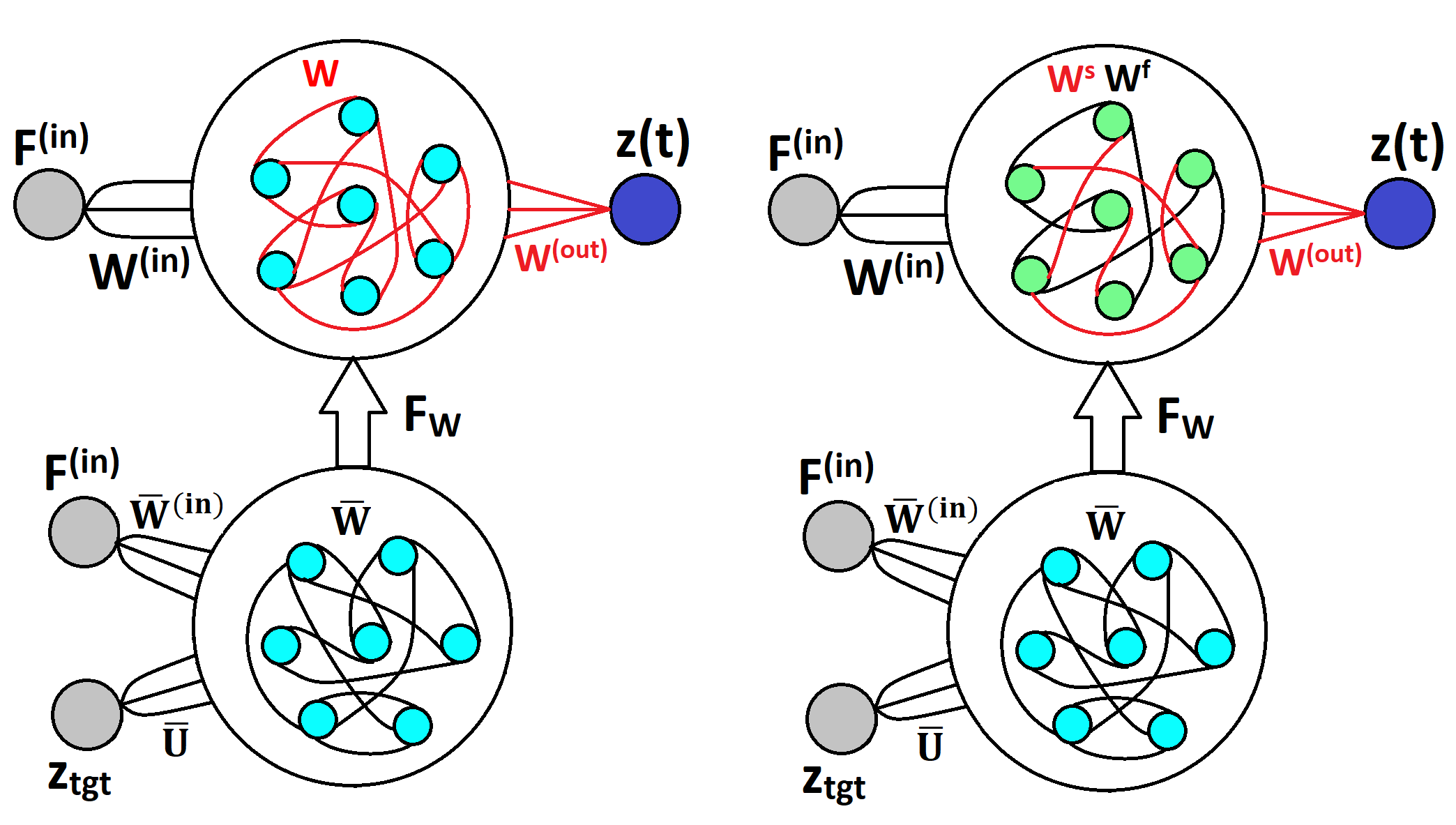}
    \caption{{\bf Full-FORCE scheme}. {\bf Left}. A rate RNN (top) is trained with the full-FORCE algorithm and supported by a target-generating network of rate neurons (bottom). {\bf Right}. A spiking RNN (top) is trained with the full-FORCE algorithm using a similar target-generating network as in the left panel. In both panels, cyan circles refer to rate neurons while green circles are LIF neurons. Dark blue circles denote the readout layer, connected to the task-performing network through plastic output weights, {\boldmath{$W^{(out)}$}}. Grey circles symbolize the input layer, which introduce a signal into the network. The task-performing networks receive the input signal, {\boldmath{$F^{(in)}$}}, but the target-generating networks also require the target signal, $z_{tgt}$, in order to generate targets specific for the task, {\boldmath{$F_W$}}. Plastic weights are symbolized by red lines while non-plastic couplings are represented by black lines. Thus, the target-generating network hidden weights ({\boldmath{$\bar{W}$}}), the input weights ({\boldmath{$W^{(in)}$}}, {\boldmath{$\bar{W}^{(in)}$}} and {\boldmath{$\bar{U}$}}) and the sRNN fast couplings ({\boldmath{$W^f$}}) are non-plastic. On the other side, the read-out synapses ({\boldmath{$W^{(out)}$}}), the rate RNN hidden couplings ({\boldmath{$W$}}) and the sRNN slow weights, {\boldmath{$W^s$}}, are plastic. Finally, $z(t)$ is the output current of the task-performing network.}
    \label{Fig:Full-FORCE}
\end{figure}

Laje and Buonomano's choice of targets is not the only possible approach. In keeping with the idea of obtaining targets from an auxiliary network, DePasquale et al. \cite{depasquale2018full} exploited the fact that the RNN in Eq. \ref{eq:dyn_rate_RNN} can solve any task and they took its recurrent currents  as targets ({\boldmath{$F_W$}}) for the corresponding currents in the RNN to be trained (Figure \ref{Fig:Full-FORCE} Left). Briefly, to obtain these targets, the target-generating network must be driven by both the desired target signals, $z_{tgt}$, and the input signals, {\boldmath{$F^{(in)}$}}. Thus, the dynamics of the target-generating network with $\bar{N}$ firing-rate units, recurrently connected through the matrix {\boldmath{$\bar{W}$}}, is governed by

\begin{equation} \label{eq:Tgt-generating_Net}
	\tau_x\frac{d\bar{x}_i(t)}{dt} = - \bar{x}_i(t) + \bar{g} \sum_{j=1}^{\bar{N}} \bar{W}_{i,j} \: r(\bar{x}_j(t)) + \sum_{l=1}^{\bar{N}_{in}} \bar{W}_{i,l}^{(in)} F_l^{(in)}(t) + \bar{U}_{i} \: z_{tgt}(t).  
\end{equation}

\noindent
To distinguish this network from the task-performing network, its synaptic weights are identified with a bar. Again, $r(\bar{x}_i)= tanh(\bar{x}_i)$ is the rate activity of unit $i$. The connection matrix, {\boldmath{$\bar{W}$}}, is built from a Gaussian distribution of zero mean and variance $1/\bar{N}$. Both the input signal, {\boldmath{$F^{(in)}$}}, and the target signal, $z_{tgt}$, are introduced through the $\bar{N}_{in}$ input synapses {\boldmath{$\bar{W}^{(in)}$}} and through the weights {\boldmath{$\bar{U}$}}, respectively. Both {\boldmath{$\bar{W}^{(in)}$}} and {\boldmath{$\bar{U}$}} are built from uniform distributions. Finally, the targets, {\boldmath{$F_W$}}, taken from the target-generating network, are given by 
	
\begin{equation}\label{eq:F_J}
	\mbox{\boldmath $F_W$}(t) = \mbox{\boldmath $u$} [g \mbox{\boldmath $\bar{W}$} \: r(\mbox{\boldmath $\bar{x}$}(t)) + \mbox{\boldmath $\bar{U}$} z_{tgt}(t) ],
\end{equation}
	
\noindent where {\boldmath{$u$}} are vectors that transfer currents in the target-generating network as targets for the task-performing network. Thus, the task-performing network weights, {\boldmath{$W$}}, are modified minimizing

\begin{equation} \label{eq:full-FORCE-1-rates}
	C_{W}^{full-FORCE} = \langle (\mbox{\boldmath $W$} \: r(\mbox{\boldmath $x$}(t)) - \mbox{\boldmath $F_W$}(t) )^2\rangle.
\end{equation}

\noindent The output weights {\boldmath{$W^{(out)}$}} are obtained as in FORCE, minimizing the cost function

\begin{equation} \label{eq:full-FORCE-2}
C_{W^{(out)}}^{full-FORCE} = \langle (z(t)-z_{tgt}(t))^2
\rangle,
\end{equation}

\noindent where $z(t)$ is the network readout current given by 

\begin{equation} 
z(t) = \mbox{\boldmath $W^{(out)}$} \: r(\mbox{\boldmath $x$}(t)).
\end{equation}

\noindent In both Eqs. \ref{eq:full-FORCE-1-rates} and \ref{eq:full-FORCE-2}, the learning rule is based on Recursive Least Squares (see Appendix \ref{app:RLS}).

\subsubsection{full-FORCE Algorithm in Spiking RNNs}

The full-FORCE algorithm can be applied to train spiking RNNs \cite{depasquale2016using} (Figure \ref{Fig:Full-FORCE}, Right). The dynamics for the sRNN is given by Eqs. \ref{eq:dyn_LIF_RNN} and \ref{eq:dyn_syn_curr}, while the dynamics for the target-generating network is Eq. \ref{eq:Tgt-generating_Net}. The targets for the hidden neurons are in Eq. \ref{eq:F_J}. The plastic connections in the task-performing network are the recurrent slow weights, {\boldmath{$W^s$}}, and the readout couplings, {\boldmath{$W^{(out)}$}} (Figure \ref{Fig:Full-FORCE}, Right). Thus, the objective is to minimize the cost of these weights. The output weights are still modified minimizing Eq.  \ref{eq:full-FORCE-2}, but with the network readout current now given by

\begin{equation} 
z(t) = \mbox{\boldmath $W^{(out)}$} \: \mbox{\boldmath $s$}(t),
\end{equation}

\noindent where {\boldmath{$s$}}$(t)$ is defined in Eq. \ref{eq:dyn_syn_curr}. The slow couplings are optimized minimizing

\begin{equation} \label{eq:full-FORCE-1-spiking}
	C_{W^s}^{full-FORCE} = \langle (\mbox{\boldmath $W^s$} \: \mbox{\boldmath $s$}(t) - \mbox{\boldmath $F_W$}(t) )^2\rangle.
\end{equation}

\subsection{Direct Transfer of Learning from rate RNNs to sRNNs}

The full-FORCE algorithm benefits from an auxiliary rate RNNs by taking the recurrent currents from the latter to use them as targets in the sRNN. Another way to leverage a rate RNN to train sRNNs is to first optimize the former and then transfer learning to the sRNN \cite{kim2019simple}. The rate RNN could be trained with any of the methods discussed above (these authors used backpropagation
through time). This is an interesting approach the implications of which have not yet been fully explored.

%%%%%%%%%%%%%%%%%%%%%%%%%%%%%
%%% SUP. LEARN. COGNITIVE %%% 
%%%%%%%%%%%%%%%%%%%%%%%%%%%%%
\subsection{RNN models of Cognitive Tasks with Supervised Learning}
\label{sect:supervised-cognitive}

Supervised learning techniques have been used to train RNNs to perform cognitive tasks, although most work was done using networks of rate units. 
As a few examples:
% russo2018motor
\cite{mante2013context,carnevale2015dynamic,song2016training,chaisangmongkon2017computing,yang2019task,wang2018flexible,dubreuil2022role}. The learning paradigm and the update rule used to train sRNNs with full-FORCE (with the learning rule RLS) or the transfer of learning from a rate RNN to a sRNN are not biologically plausible. However, the trained networks do comply with two important biological features (network connectivity and neuron type) and both the network performance and neural activity exhibit many features found in behaving animals. This was recently seen in sRNNs trained to a delayed match-to-sample task \cite{kim2021strong} and in sRNNs trained to discriminate on the duration of temporal intervals \cite{serrano2022TID}.

%%%%%%%%%%%%%%%%%%%%%%%%%%%%%%%
%%%% REINFORCMENT LEARNING %%%% 
%%%%%%%%%%%%%%%%%%%%%%%%%%%%%%%
\section{Reinforcement Learning}
\label{sect:RL}
The need of a teacher limits the use of supervised methods to cases where a teaching signal is supplied; instead, Reinforcement Learning (RL) uses a numerical \textit{reward} as a teaching signal which, for the tasks we are concerned with, is more realistic. Although the normative basis were set almost 40 years ago and recently compiled by their pioneers \cite{sutton2018reinforcement}, here we first discuss the key concepts that will allow us to introduce in Sect. \ref{subsect:RNN_implementation_RL} their implementation in the RNN framework. A more formal description of the RL framework is summarized in Appendix \ref{app:rl}.

\subsection{Introduction to Reinforcement Learning} 
In RL the goal of an \textit{agent} is to learn to map \textit{states} $s \in \mathcal{S}$ of an \textit{environment} to certain \textit{actions} $a \in \mathcal{A}(s)$ that maximize a numerical reward $r \in \mathcal{R}$. Thus, this reward acts as a teaching signal. As opposed to supervised learning, this signal is not unambiguously given to the  agent, but it is rather discovered by the agent through  his \textit{exploration}  strategies of the environment. The learning process is considered complete once the agent correctly \textit{exploits} the optimized mapping, hence maximizing, on average, the aforementioned reward. 

\paragraph{Return and cumulative reward functions.} The agent's goal is to maximize the average of a \textit{return} function $G_t$ at each time step $t$. This function considers not only the immediate reward $R_{t+1}$, but also the total amount of rewards the agent encounters along a trajectory. We can formalize this idea with the following definition,

\begin{equation}
    \label{eq:return}
    G_t \doteq R_{t+1} + \gamma R_{t+2} + \gamma^2 R_{t+3} + ... = \sum_{\tau=t+1}^{T} \gamma^{\tau-t-1} R_{\tau},
\end{equation}

\noindent where $T$ is the last step of the episode (trial) and $\gamma$ is a parameter, $0 \leq \gamma \leq 1$, called the \textit{discount} factor or rate. Grossly, $\gamma$ accounts for the fact that a reward received far in the future has less value than a reward received immediately.

\paragraph{Policies and value functions.} \label{par:Pol-value} The agent interacts with the environment by means of a \textit{policy} $\pi(a|s)$ that map states to available actions. At each time step $t$ an action $A_t \sim \pi(a|S_t=s)$ is sampled from the policy. To evaluate the action suggested by the current policy, the agent computes a \textit{state-value} function $\upsilon_{\pi}(S_t=s)$ dependent only on the state, or an \textit{action-state} value function $q_{\pi}(A_t=a,S_t=s)$ that incorporates the fact that, in a certain state, different actions might lead to different consequences. Briefly, value functions store \textit{predictions} of future rewards that are expected if the agent follows a certain policy, hence addressing a temporal version of the CAP (see Sect. \ref{subsubsect:CAP}). 
However, storing discrete predictions for each state and action is not feasible for complex environments where the state space is impractically large \cite{moore1990efficient} or when the agent has access to noisy features (e.g., probabilities) of those states \cite{kaelbling1998planning,rao2010decision,sarno2017dopamine,sarno2022dopamine}. It is then convenient to approximate $\upsilon_{\pi}(s)$ by means of a parameterized function $\nu_{\pi}(s,\bm{\phi})$ with $\bm{\phi}$ being a weight vector of arbitrary dimensions\footnote{If necessary, the same applies to state-action values $q_{\pi}(s,a)$.}. Similarly, the policy can also be expressed as a functional form $\pi(a|s,\mbox{\boldmath{$W$}})$ of some parameters $\mbox{\boldmath{$W$}}$.

Importantly, when implementing RL in neural networks, the parameters $\bm{\phi}$ and $\mbox{\boldmath{$W$}}$ are taken as plastic synaptic couplings of the corresponding neural networks. By parameterizing policies and value functions, the RL problem is transformed into a weight optimization problem. Policy optimization is fundamentally different from value function approximation. The former \textit{controls} the agent's actions so rewards are maximized, while the latter seeks to accurately \textit{predict} them. 

\paragraph{Policy Optimization and REINFORCE.} \label{par:PG-REINFORCE} If the goal is to maximize an objective function given by the average discounted total reward, we can modify the policy $\pi$ by updating the parameters $\mbox{\boldmath{$W$}}$ equating their change to the gradient of the state-value fucntion $\upsilon_{\pi}(s)$ (see Eq. \ref{eq:state-value}). 

Its gradient with respect to the parameters $\mbox{\boldmath{$W$}}$ is given by the Policy Gradient Theorem\footnote{Note that we abuse notation by dropping the subscript $\mbox{\boldmath{$W$}}$ in the gradient  (i.e., $\nabla_{\mbox{\boldmath{$W$}}} \longrightarrow \nabla$).} \cite{sutton1999policy}:

\begin{equation}
    \label{eq:policy-gradient}
    \begin{split}
    \nabla \upsilon_{\pi}(s) & \propto \sum_{s \in \mathcal{S}} \mu_{\pi} (s) \sum_{a \in \mathcal{A}} q_{\pi}(s,a) \nabla \pi (a|s,\mbox{\boldmath{$W$}}) \\
& \propto \sum_{s \in \mathcal{S}} \mu_{\pi} (s) \sum_{a \in \mathcal{A}} \Big( q_{\pi}(s,a) - b(s) \Big) \nabla \pi (a|s,\mbox{\boldmath{$W$}}) \\
& = \sum_{s \in \mathcal{S}} \mu_{\pi} (s) \sum_{a \in \mathcal{A}} \pi (a|s,\mbox{\boldmath{$W$}})\Big( q_{\pi}(s,a) - b(s) \Big) \frac{\nabla \pi (a|s,\mbox{\boldmath{$W$}})}{\pi (a|s,\mbox{\boldmath{$W$}})} \\
& = \mathbb{E}_{\pi} \left[ \Big( G_t - b(S_t=s) \Big) \nabla \text{ln } \pi (A_t=a|S_t=s,\mbox{\boldmath{$W$}}) \right]. \\
    \end{split}
\end{equation}

\noindent  The baseline $ b(s)$ added in the second line is arbitrary, since the sum over actions is zero if the policy is normalized \cite{sutton2018reinforcement}. We will see its purpose below. A key operation to obtain a REINFORCE type algorithm \cite{williams1992simple}, is the replacement of the sum over random state-action pairs $(s,a)$ by their sampling (fourth line). Additionally, $q_{\pi}(s,a)$ was replaced by $G_t$ (see Eq. \ref{eq:action-value}). Provided that the sampling is done following policy $\pi$, the quantity inside the square brackets is a reliable estimate of $\nabla \upsilon_{\pi}(s)$ \cite{sutton1999policy}.  The factor $\nabla \text{ln } \pi (A_t|S_t,\mbox{\boldmath{$W$}})$, known as \textit{eligibility trace}, follows from the identity $\partial \text{ln }f = f^{-1}\partial f$. 

The return $G_t$ in Eq. \ref{eq:policy-gradient} is only accessible after a whole episode has been sampled. By averaging over an entire episode of duration $T$, a single update of the parameters at the end of the episode is possible, 

\begin{equation}
    \label{eq:unique-update-G}
    \Delta \mbox{\boldmath {$W$}} = \eta \sum_{t=0}^{T} \Big( G_t-b(S_t) \Big) \nabla \text{ln } \pi (A_t|S_t,\mbox{\boldmath {$W$}}),
\end{equation}

\noindent  where the learning rate $\eta$ can accommodate the normalizing factor $1/T$. Note that averaging over an entire episode, or even across several trials, is formally identical to accumulating the gradient step over each time and trial and then doing the gradient step. Furthermore, if a single reward is delivered at the terminal state $T$, no discount is considered (see Eq. \ref{eq:return-simple}), and if the baseline is state-independent, the REINFORCE update simplifies to 

\begin{equation}
    \label{eq:unique-update-R}
    \Delta \mbox{\boldmath {$W$}} = \eta (R_T-b) \sum_{t=0}^{T} \nabla_{\mbox{\boldmath {$W$}}} \text{ln } \pi (A_t|S_t,\mbox{\boldmath {$W$}}).
\end{equation} 

\noindent As mentioned above, the choice of baseline is arbitrary. However, this baseline affects the convergence of the algorithm. For example, it can be used to control the fluctuations of the gradient estimator, reducing its variance (see, e.g., \cite{fremaux2010functional}).

\paragraph{Predictions and Actor-Critic Architectures.} \label{par:AC} As opposed to Policy Gradient methods, which optimize the action selection process to maximize the average return, value based methods aim to find the best estimation of it. The parameters $ \bm{\phi}$ can be optimized by performing stochastic gradient descent on the Mean Squared Error between the return $G_t$ and the parameterized function $v_{\pi}(s,\bm{\phi})$

\begin{equation} \label{eq:MSE-value}
    E(\bm{\phi}) = \mathbb{E}_{s\sim \pi} \Big[ \frac{1}{2} \Big( G_t-\nu_{\pi}(s,\bm{\phi})\Big)^2 \Big],
\end{equation}

\noindent where the expectation is respect to sampled states following policy $\pi$. The difference inside the parenthesis, which is formally identical to the factor appearing in the REINFORCE estimate (e.g., Eq. \ref{eq:unique-update-G}), is a form of \textit{Prediction Error}. Although other choices are possible (see Section \ref{subsect:RNN_implementation_RL}), value functions are a natural choice for the baseline $b(s)$ giving rise to a whole family of Actor-Critic architectures \cite{sutton2018reinforcement}. In these architectures, an \textit{actor} is in charge of selecting actions according to a policy $\pi$ while a \textit{critic} computes the Prediction Error. Then, the policy is updated according to this criticism. Interestingly, certain neurons in the mammalian brain are specialized in the computation of prediction errors to guide learning \cite{schultz2000neuronal}, possibly of Hebbian type \cite{roelfsema2018control}.

\subsection{RNN implementations of Reinforcement Learning}
\label{subsect:RNN_implementation_RL}

So far, in the search for biologically realistic training methods, we stressed those dealing with recurrent networks of spiking neurons, without worrying about the biological plausibility of the learning paradigm or the learning rule. More realistic training methods should employ the RL paradigm, instead of supervised learning. However, although there have been some proposals to train RNNs with RL methods, they often deal with continuous-variable neurons \cite{miconi2017biologically,song2017reward}. Another concern is that the learning rule in Eq.  \ref{eq:unique-update-G} (or Eq. \ref{eq:unique-update-R}) is not biologically plausible. Unfortunately, there are still no widely accepted methods that train RNNs of spiking neurons within RL paradigm and use realistic learning rules. Available RL methods often ignore the spiking neurons, or the plausibility of the learning rule, or relevant biological features of the network architecture. Even so, there are some notable works that could indicate a way forward in finding better models of RNNs. We next refer to two of them  \cite{miconi2017biologically,song2017reward}. For other approach and network architecture see \cite{vasilaki2009spike}.

\subsubsection{A Biologically Plausible Form for the Eligibility Trace}
\label{Miconi_2017}

Miconi's work \cite{miconi2017biologically} is focused on training RNNs of rate neurons within the RL paradigm and a biologically realistic learning rule. For this purpose, he trains RNNs of rate neurons (as in Eq. \ref{eq:dyn_rate_RNN}), but adds a stochastic perturbation $\Delta_i(t)$ to the neuron's activity

\begin{equation}
	\tau_x\frac{dx_i(t)}{dt} = - x_i(t) +  g \sum_{j=1}^{N} W_{i,j} r_j + \sum_{l=1}^{N_{in}} W_{i,l}^{(in)} F_l^{(in)}(t) + \Delta_i(t),  \label{eq:dyn_rate_RNN_pertub}
\end{equation}

\noindent  The perturbations $\Delta_i(t)$ are spatially and temporally uncorrelated (perturbations are applied in both learning and testing stages). To devise a plausible learning rule, he builds on  earlier proposals \cite{fiete2006gradient,hoerzer2014emergence,legenstein2010reward}, arriving to a three-factor Hebbian-like rule modulated by a reward factor

\begin{equation} \label{eq_Miconi_update}
\Delta W_{i,j} = \eta (R_T - \bar{R}) e_{i,j},
\end{equation}

\noindent  where $R_T$ denotes a single reward delivered at the end of the trial (See Eq. \ref{eq:return-simple}) and $\bar{R}$ is the average reward obtained by the agent (used here as a reward baseline). It is computed as a running average of recent rewards for trials of the same type. If $R(n)$ is the reward for trial $n$, $\bar{R}$ is updated as

\begin{equation} \label{eq_Miconi_R_running_avg}
\bar{R}(n) = \alpha \bar{R}(n-1) +  (1 - \alpha) {R}(n).
\end{equation}

\noindent  Note that the second factor in Eq. \ref{eq_Miconi_update} is a reward prediction error. The eligibility trace $e_{i,j}$, accumulates Hebbian-like products of input and output signals (see Eq. \ref{eq_Hebbian_Rule}),

\begin{equation} \label{eq_Miconi_EETT}
e_{i,j}(t) = e_{i,j}(t-1)  + S \Big( r_j(t-1) (x_i(t) - \bar{x_i}) \Big).
\end{equation}

\noindent  Here $\bar x_i$ denotes a short-term running average of $x_i$, then the difference in the last factor tracks activity fluctuations. $S$ is a monotonic, supralinear function of its inputs (e.g., $x^3$) which implies that the plasticity mechanism is dominated by large increments, and tends to suppress smaller ones. The 
 equations \ref{eq_Miconi_update}-\ref{eq_Miconi_EETT} have the form of the update in the REINFORCE algorithm, Eq. (\ref{eq:unique-update-R}), with the important addition that the accumulation present in the eligibility trace is inspired by biologically plausible arguments. Although there is no mathematical proof that the learning rule actually approximates the policy gradient, in practice it works well (see Section \ref{subsect:RL-cognitive_tasks}).

\subsubsection{A RNN that computes the reward baseline}
\label{Song_Wang_2017}

Note that Miconi's training method does not provide a neural implementation to retrieve the reward baseline $\bar{R}$, but it is computed algorithmically with Eq. \ref{eq_Miconi_R_running_avg}. A proposal to estimate a baseline by means of a second RNN was given by Song et al. \cite{song2017reward} (but see also \cite{wierstra2009recurrent}). 
Their network model is composed of two modules arranged in an Actor-Critic architecture (see Section \ref{par:AC}), in which one of the RNNs (the decision or control network) selects actions that maximize reward, while the other RNN (the prediction network, also named the value network) uses the selected actions and the activity of the decision network to predict future reward and compute the baseline $b$, which es presented as the output of this network.

Both the decision network and the value network are rate RNNs, with synaptic weights ${\mbox{\boldmath{$W$}}}$ and $\bm{\phi}$, respectively. The activities of unit $i$ in each network (respectively $x_i^{\pi}$ and  $x_i^{\nu}$) follow equations similar to those presented in Eq. (\ref{eq:dyn_rate_RNN}). However, Song et al. consider more complex rate units than those previously discussed in Eq. (\ref{eq:dyn_rate_RNN}), called Gated Recurrent Units \cite{chung2014empirical}. %\cite{cho2014learning,chung2014empirical}. 
Another difference is the  contribution of Gaussian independent noises $\xi_i$, with variance $\sigma$, of the form $\sqrt{2 \tau_x \sigma^2 \xi_i}$. The firing rates, $r_i^{\pi}$ and  $r_i^{\nu}$, are related to the corresponding activities by linear-threshold functions. 

In supervised learning, the activity of the output neurons is used as a decision variable. Instead, these two networks use output units for the following purposes: First, the output of the decision network determines the numerical preference at time $t$ ($z_i(t)$) for the action $k$ ($k=1, \ldots, N_a$),

\begin{equation} \label{eq_decision_net_out}
z_k(t)=\sum_{j=1}^{N} \Big( W_{k,j}^{(\pi,out)} r_j^{\pi}(t) + b_k^{\pi,out} \Big).
\end{equation}

\noindent  At each time step, actions are randomly sampled from the corresponding policy parameterized according to a \textit{softmax} distribution (Appendix \ref{app:rl-policies}), 

\begin{equation} \label{eq_policy}
\pi(a_t=k|\mbox{\boldmath {$S$}}_{1:t},\mbox{\boldmath{$W$}}) = \frac{e^{z_k(t)}}{\sum_{l=1}^{N_a}e^{z_l(t)}}. 
\end{equation}

\noindent  Second, the single output of the value network delivers a predictions for the future return, 

\begin{equation} \label{eq_prediction_net_out}
\nu_{\bm{\phi}}(\mbox{\boldmath {$a$}}_{1:t},\mbox{\boldmath {$r$}}^{\pi}_{1:t}) = \sum_{j=1}^{N} \Big( W_{j}^{(\nu,out)} r_j^{\nu}(t) + b^{\nu,out} \Big).
\end{equation}

\noindent  The two networks cooperate to learn an optimal strategy. The decision network receives the stimuli as inputs  (and any other necessary information about the environment such as cues and context) and delivers the policy in its output (Eqs. \ref{eq_decision_net_out}-\ref{eq_policy}). The agent, at each trial time, uses the policy to select an action. The environment receives the selected action and responds by making a transition to another state and, depending on the selected action, delivering a reward. Furthermore, the decision network also sends the selected action to the prediction network, along with the activities of its units. Finally, the value network uses this data to predict the expected return at time t, which it displays in its output (Eq. \ref{eq_prediction_net_out}).

As in Miconi 2017, in this paper, the agent's goal is to maximize the sum of expected future rewards. This is done in the decision network, which is trained with the REINFORCE algorithm \cite{williams1992simple}. Also as in \cite{miconi2017biologically}, the synaptic couplings update is done according to REINFORCE, however, unlike that work, instead of searching for a biologically realistic way to estimate the eligibility trace, Song et al. estimate it as given by the policy gradient theorem, using Monte Carlo. More precisely, the update of the parameters (synaptic couplings of the decision network) in Eq. \ref{eq:unique-update-G} is written as

\begin{equation}
\label{eq_REINFORCE_update_Song}
\Delta W_{i,j} = \eta \sum_{t=0}^{T}\Big (G_t-\nu_{\bm{\phi}}(\mbox{\boldmath {$a$}}_{1:t},\mbox{\boldmath {$r$}}^{\pi}_{1:t}) \Big)  \frac{\partial \text{ln } \pi (a_t|\mbox{\boldmath {$S$}}_{1:t},\mbox{\boldmath {$W$}})}{\partial W_{i,j}} , 
\end{equation}

\noindent with the baseline $b$ given by the output $\nu_{\bm{\phi}}$ of the prediction network. In turn, the value network is optimized by minimizing the prediction error $G_t - \nu_{\bm{\phi}}(\mbox{\boldmath {$a$}}_{1:t},\mbox{\boldmath {$r$}}^{\pi}_{1:t})$ from Eq. \ref{eq:MSE-value}.

%%%%%%%%%%%%%%%%%%%%%%%%%%%%%%%%
%%%%%%%%% RL COGNITIVE %%%%%%%%% 
%%%%%%%%%%%%%%%%%%%%%%%%%%%%%%%%
\subsection{RNN models of Cognitive Tasks with Reinforcement Learning}
\label{subsect:RL-cognitive_tasks}

Miconi applied his method to successfully solve two cognitive tasks \cite{miconi2017biologically}: a delayed XOR problem (delayed nonmatch-to-sample task) and a context-dependent sensory integration task \cite{mante2013context}. The sensory integration task required  attending selectively to specific aspects of the sensory input, while ignoring the others, in a variable context.

To analyse the performance of their training procedure, Song et al. trained networks to solve several popular cognitive scenarios \cite{song2017reward}: a random dots motion task \cite{kiani2008bounded}, the same context-dependent sensory integration task as Miconi \cite{mante2013context}, a multisensory integration task \cite{raposo2012multisensory}, a parametric working memory task \cite{romo1999neuronal}, a postdecision wager task \cite{kiani2009representation}, where the optimal decision depends on the agent's internal state and a value-based economic task, to examine the activity of the value rather than the the decision network \cite{padoa2006neurons}. In all cases the trained networks exhibited properties similar to those observed in  experiments with behaving animals.

%%%%%%%%%%%%%%%%%%%%%%%%%%%%
%%%% BIOLOGICAL PLAUS %%%%%% 
%%%%%%%%%%%%%%%%%%%%%%%%%%%%
\section{Discussion on Biological Plausibility}
\label{sect:biological}

To obtain more biologically plausible models of credit assignment, researchers have proceeded in several directions. One approach attempts to bring together the most salient biological features, e.g., spiking neurons and recurrent connectivity \cite{depasquale2016using, kim2018learning,kim2019simple,nicola2017supervised, kim2021strong, bellec2020solution, serrano2022TID}.  
However, most of these studies neglect the plausibility of the learning paradigm and the learning rule. For example, recursive least squares, used by \cite{depasquale2016using, kim2018learning,serrano2022TID} to train sRNNs, is not a local rule, as required by biology. Also backpropagation-through-time, used in \cite{kim2019simple, kim2021strong}, lacks biological realism  \cite{lillicrap2019backpropagation}. In addition, none of those works trains sRNNs with reward-based algorithms. 
Other authors trained RNNs focusing on reinforcement learning methods \cite{miconi2017biologically, song2017reward} and plausible learning rules \cite{miconi2017biologically}, while neglecting the biological realism of the neuron model. Another aspect of biological neural networks is that they consist of several neural circuits, both during learning and task execution. While the work of Song et al. is entirely based on non-realistic ML techniques, it has the particularity of using two networks for training with RL methods, which has a possible correspondence with brain anatomy.
In short, there is currently no credit assignment model that simultaneously considers sRNNs, trained with reward-based algorithms and biological plausible learning rules. Note, however, that the optimized sRNNs do bring together two essential properties of cortical networks, and can be used to investigate how they solve cognitive tasks \cite{ kim2021strong,serrano2022TID}.

Because error backpropagation has computational requirements that are difficult to match with neural mechanisms, it was initially thought that it could not be implemented by the brain (see \cite{lillicrap2020backpropagation} and references therein). However, several efforts have been made recently to approximate it with biological properties of neurons and cortical circuits \cite{ guerguiev2017towards, sacramento2018dendritic,richards2019dendritic,payeur2021burst}. These approaches consider that neurons are not so simple as LIF neurons. The most abundant cortical neurons are pyramidal cells, which posses three basic dendritic types (see \cite{richards2019dendritic} and references therein). Most of the cell's  feedforward and recurrent inputs reach the basal dendrites and soma dendrites, but a third dendrite type (distal apical dendrites), which is electrotonically distant from the others, receive feedback signals from higher-order brain areas. This is a key feature of pyramidal cells, since these distal dendrites may provide a route for credit assignment that takes care of feedback signals. In this way, the credit assignment computations instructed by signals coming from higher-order areas can be done independently from the sensory signals arriving from the somatic dendrites. Several proposals to introduce credit assignment from feedback signals are based on this mechanism (e.g.,  \cite{kording2001supervised, guerguiev2017towards, sacramento2018dendritic}). Other authors sought to avoid backpropagation through hidden layers or through time by training only the output and the top-down weights with RL \cite{Masse2022.05.09.491102}.

In this paper we gave an overview of the state-of-the-art on how to train neural networks with biological realism, focusing on some key contributions that could serve as a basis for future efforts to advance this field of research.  
Although there are no widely accepted methods for training neural networks with biological plausibility, those contributions have begun to point a way forward.

We stressed the need to train networks of spiking neurons, with recurrent architecture, using biologically-based learning paradigms and rules. 
The success achieved in understanding the network mechanisms at work in various cognitive tasks  \cite{mante2013context,barak2013fixed, carnevale2015dynamic,song2016training,miconi2017biologically,song2017reward, chaisangmongkon2017computing,wang2018flexible,yang2019task,kim2021strong, serrano2022TID} encourages revisiting classic RNN models proposed earlier (e.g., \cite{parga1986ultrametric,renart1999backward,renart2001model,brunel2001effects}). Several significant aspects have been left out of this account. For instance, although most current work trains a single neural network, in the brain cognitive tasks are solved in a distributed manner by sets of cortical circuits. Also, a single cost function is usually used, whereas different brain areas may have different objectives. Moreover, cortical circuits are not restricted to participate in a single task. Although more rarely, some of these issues are being addressed \cite{yang2021towards,yang2019task}. The prospects for progress in modeling cognitive tasks with biologically realistic networks appear to be promising.

\appendix
%%%%%%%%%%%%%%%%%%%%%%%%%%%%%%%%
%%%%%%%%%% RLS %%%%%%%%%%
%%%%%%%%%%%%%%%%%%%%%%%%%%%%%%%%
\section{Recursive Least Squares Algorithm}
\label{app:RLS}

Recursive Least Square (RLS) \cite{sussillo2009generating} is used to minimize any cost function. This algorithm looks recursively for the least square error, with error defined as the difference between the desired signal, $z_{tgt}(t)$, and the output signal, $z(t)$: $error(t) = z(t) - z_{tgt}(t).$ Depending on the model, $z(t)$ could be obtained from

\begin{equation*}
	z(t) = \mbox{\boldmath $W$} \: r(\mbox{\boldmath $x$}(t)), \label{eq:App_RLS_2}
\end{equation*} 

\noindent in rate RNNs, or from slow synaptic currents {\boldmath{$s$}}(t) in (Eq. \ref{eq:dyn_syn_curr})

\begin{equation*}
	z(t) = \mbox{\boldmath $W$} \: \mbox{\boldmath $s$}(t)). \label{eq:App_RLS_3}
\end{equation*}

\noindent From now on, we continue for the case of spiking neurons, but the case of rate neurons is done in an analogous way. The plastic synapses, {\boldmath{$W$}}, are updated as follows for rate neurons

\begin{equation}
	\mbox{\boldmath $W$}(t) = \mbox{\boldmath $W$}(t-\Delta t) - error(t) \: \mbox{\boldmath $P$}(t) \: \mbox{\boldmath $s$}(t). \label{eq:App_RLS_4}
\end{equation} 

\noindent 
Here {\boldmath{$P$}}(t) is an estimation of the correlation matrix of {\boldmath{$s$}}(t) and its  upgrade is made according to
	
\begin{equation}\label{eq:App_RLS_P}
	\mbox{\boldmath $P$}(t) = \mbox{\boldmath $P$}(t-\Delta t) - \frac{\mbox{\boldmath $P$}(t-\Delta t) \: \mbox{\boldmath $s$}(t) \: \mbox{\boldmath $s$}(t)^T \: \mbox{\boldmath $P$}(t - \Delta t)}{1 + \mbox{\boldmath $s$}(t)^T \: \mbox{\boldmath $P$}(t-\Delta t) \: \mbox{\boldmath $s$}(t)}. 
\end{equation}

%%%%%%%%%%%%%%%%%%%%%%%%%
%%%%%% RL EXTRAS %%%%%%%% 
%%%%%%%%%%%%%%%%%%%%%%%%%

\section{Mathematical Considerations of Reinforcement Learning}
\label{app:rl}

\paragraph{\em Markov Decision Processes.}
\label{app:rl-MDPs}
A more formal description of RL involves a Markov Decision Process (MDP). Formally, a MDP is an idealized mathematical representation uniquely defined by a set of states $\mathcal{S}$ and actions in each state $\mathcal{A}(s)$, a reward function $\mathcal{R} \in \mathbb{R}$ and an environment whose dynamics $\mathcal{T}$ are, in general, stochastic (see below). We next clarify what is meant by environment and agent. An environment is a mathematical abstraction containing a set of states $\mathcal{S}$ that can be visited sequentially in a discrete set of time steps, $t=0,1,2,...$. When the agent is in the state $S_t \in \mathcal{S}$ it chooses an action $A_t \in \mathcal{A}(s)$, then the environment transitions with a certain probability $\mathcal{T}$ to a new state $S_{t+1} \in \mathcal{S}$ obtaining a reward $R_{t+1} \in \mathcal{R} \subset \mathbb{R}$. As a result of this iterative process, the interaction between the agent and the environment gives rise to a \textit{trajectory} (i.e., a \textit{history}) $\mathcal{H}=S_0, A_0, R_1, S_1, A_1, R_2, S_2, A_2, \ldots$ . 

\paragraph{\em Insigths on the return function.} 
\label{app:rl-return}
In the limit case where $\gamma = 0$, the return $G_t$ in Eq. \ref{eq:return} is equivalent to the immediate reward $G_t = R_{t+1}$. In contrast, in the absence of any discount, the return $G_t$ is identical to the \textit{cumulative reward} function $G_t=\sum_{\tau=t+1}^{T} R_{\tau}$. Therefore, in the general case, the goal of a RL agent is to maximize the expectancy of the discounted cumulative reward function. A particularly useful simplification of the return function arises when dealing with sparse rewards delivered only at the end of the episodes. From Eq. \ref{eq:return} it is straightforward to see that if all rewards are 0 for non-terminal states the return function is written as

\begin{equation} \label{eq:return-simple}
G_t = \gamma^{T-t-1} R_T \stackrel{\gamma=1}{\longrightarrow} R_T.
\end{equation}

\paragraph{\em Stochasic Policies.}
\label{app:rl-policies} 

As stated in section \ref{sect:RL}, a policy is a function that assigns which actions are available in a given state. Once an agent is in a given state $s$, it samples an action $a$ from the corresponding policy $\pi(a|s)$. Deterministic policies ensure that, provided the agent is in a specific state $\tilde{s}$, the same action $\tilde{a}$ is always sampled.

Deterministic policies are interesting provided they are optimal, but it is worth noting that optimal policies can be stochastic (i.e., non-deterministic). Even more, if we consider the whole learning process, agents often start with a complete random policy and they need to sample or explore actions while at the same time exploiting those that have been proven to deliver high rewards. Indeed, how much to explore and exploit is a non-trivial dilemma that can be solved with several policy schemes. Nonetheless, one of the most common policy parameterizations, \textit{softmax}, consists in an exponential weighting of a set of \textit{preferences} for each state-action pair $ h(s,a,\mbox{\boldmath{$W$}}) \in \mathbb{R}$ where, once again, $\mbox{\boldmath{$W$}}$ is some weight vector of arbitrary dimensions. 

\begin{equation}
    \label{eq:softmax}
    \pi(a|s,\mbox{\boldmath{$W$}}) = \frac{e^{h(s,a,\mbox{\boldmath{$W$}}})}{\sum_{k\in\mathcal{A}(s)}e^{h(s,k,\mbox{\boldmath{$W$}})}},
\end{equation}

\noindent where $e \approx 2.718$ is the base of the natural logarithm. Note that the denominator is a constant ensuring that the policy $\pi$ is normalized to unity. The numerical preferences control the selection process, that is, actions with higher preference will be exponentially more probable. Preferences are usually initialized using random weights $\mbox{\boldmath{$W$}}$ thus resulting in fully exploratory behaviors. As rewards are encountered and preferences are accordingly updated (e.g., as in Eq. \ref{eq:unique-update-G}), the \textit{softmax} schema gradually converges to a policy that guarantees an optimal exploitation of the environment.

\paragraph{\em Value functions.}
\label{app:rl-value} 

The state-value function is formally defined as
\begin{equation}
    \label{eq:state-value}
    \upsilon_{\pi}(s) \doteq \mathbb{E}_{\pi}[G_t | S_t=s] \text{ ,   } \forall s \in \mathcal{S}
\end{equation}

\noindent and the action-value function as

\begin{equation}
    \label{eq:action-value}
    q_{\pi}(s,a) \doteq \mathbb{E}_{\pi} [G_t | S_t=s, A_t=a] \text{ ,   } \forall s \in \mathcal{S} \text{ and } \forall a \in \mathcal{A}(s)
\end{equation}

\noindent where $\mathbb{E}_{\pi}[\cdot]$ denotes the expected value following a given policy. The subscript $\pi$ is necessary due to the inherent dependency of the return $G_t$ on the policy followed to sample a given trajectory $\mathcal{H}$. Although similar, the $q_{\pi}$ function captures the fact that different actions might carry different rewards at any particular state. In value-based RL approaches, the action-value function is considered as the preference $h(\cdot)$ in Eq. \ref{eq:softmax}, hence the policy is updated according to the predictions of rewards.

%%%%%%%%%%%%%%%%%%%%%%%%%%%%%%%%
%%%%%%% ACKNOWLEDGEMENTS %%%%%%% 
%%%%%%%%%%%%%%%%%%%%%%%%%%%%%%%%
\section*{Acknowledgements}
This research was supported by PGC2018-101992-B-I00 from the Spanish Ministry of
Science, Innovation and Universities (to L.S.-F. and N.P.), by European Union’s Horizon 2020 research and innovation program under grant agreement Sano No 857533 and by Sano project carried out within the International Research Agendas program of the Foundation for Polish Science, co-financed by the European Union under the European Regional Development Fund (to J.F.-R.).

%%%%%%%%%%%%%%%%%%%%%%%%%%%%%%%%
%%%%%%%%%% REFERENCES %%%%%%%%%% 
%%%%%%%%%%%%%%%%%%%%%%%%%%%%%%%%
\bibliographystyle{ieeetr}
\bibliography{references}  

\begin{thebibliography}{10}

\bibitem{sussillo2009generating}
D.~Sussillo and L.~F. Abbott, ``Generating coherent patterns of activity from
  chaotic neural networks,'' {\em Neuron}, vol.~63, no.~4, pp.~544--557, 2009.

\bibitem{laje2013robust}
R.~Laje and D.~V. Buonomano, ``Robust timing and motor patterns by taming chaos
  in recurrent neural networks,'' {\em Nature neuroscience}, vol.~16, no.~7,
  pp.~925--933, 2013.

\bibitem{depasquale2016using}
B.~DePasquale, M.~M. Churchland, and L.~Abbott, ``Using firing-rate dynamics to
  train recurrent networks of spiking model neurons,'' {\em arXiv preprint
  arXiv:1601.07620}, 2016.

\bibitem{depasquale2018full}
B.~DePasquale, C.~J. Cueva, K.~Rajan, G.~S. Escola, and L.~F. Abbott,
  ``full-force: A target-based method for training recurrent networks,'' {\em
  PLOS ONE}, vol.~13, pp.~1--18, 02 2018.

\bibitem{kim2019simple}
R.~Kim, Y.~Li, and T.~J. Sejnowski, ``Simple framework for constructing
  functional spiking recurrent neural networks,'' {\em Proceedings of the
  national academy of sciences}, vol.~116, no.~45, pp.~22811--22820, 2019.

\bibitem{miconi2017biologically}
T.~Miconi, ``Biologically plausible learning in recurrent neural networks
  reproduces neural dynamics observed during cognitive tasks,'' {\em Elife},
  vol.~6, p.~e20899, 2017.

\bibitem{song2017reward}
H.~F. Song, G.~R. Yang, and X.-J. Wang, ``Reward-based training of recurrent
  neural networks for cognitive and value-based tasks,'' {\em Elife}, vol.~6,
  p.~e21492, 2017.

\bibitem{romo1999neuronal}
R.~Romo, C.~D. Brody, A.~Hern{\'a}ndez, and L.~Lemus, ``Neuronal correlates of
  parametric working memory in the prefrontal cortex,'' {\em Nature}, vol.~399,
  no.~6735, pp.~470--473, 1999.

\bibitem{padoa2006neurons}
C.~Padoa-Schioppa and J.~A. Assad, ``Neurons in the orbitofrontal cortex encode
  economic value,'' {\em Nature}, vol.~441, no.~7090, pp.~223--226, 2006.

\bibitem{kiani2009representation}
R.~Kiani and M.~N. Shadlen, ``Representation of confidence associated with a
  decision by neurons in the parietal cortex,'' {\em science}, vol.~324,
  no.~5928, pp.~759--764, 2009.

\bibitem{raposo2012multisensory}
D.~Raposo, J.~P. Sheppard, P.~R. Schrater, and A.~K. Churchland, ``Multisensory
  decision-making in rats and humans,'' {\em Journal of neuroscience}, vol.~32,
  no.~11, pp.~3726--3735, 2012.

\bibitem{jazayeri2015neural}
M.~Jazayeri and M.~N. Shadlen, ``A neural mechanism for sensing and reproducing
  a time interval,'' {\em Current Biology}, vol.~25, no.~20, pp.~2599--2609,
  2015.

\bibitem{wang2018flexible}
J.~Wang, D.~Narain, E.~A. Hosseini, and M.~Jazayeri, ``Flexible timing by
  temporal scaling of cortical responses,'' {\em Nature neuroscience}, vol.~21,
  no.~1, pp.~102--110, 2018.

\bibitem{de2005neuronal}
V.~de~Lafuente and R.~Romo, ``Neuronal correlates of subjective sensory
  experience,'' {\em Nature neuroscience}, vol.~8, no.~12, pp.~1698--1703,
  2005.

\bibitem{carnevale2013optimal}
F.~Carnevale, V.~de~Lafuente, R.~Romo, and N.~Parga, ``An optimal decision
  population code that accounts for correlated variability unambiguously
  predicts a subject’s choice,'' {\em Neuron}, vol.~80, no.~6,
  pp.~1532--1543, 2013.

\bibitem{mante2013context}
V.~Mante, D.~Sussillo, K.~V. Shenoy, and W.~T. Newsome, ``Context-dependent
  computation by recurrent dynamics in prefrontal cortex,'' {\em nature},
  vol.~503, no.~7474, pp.~78--84, 2013.

\bibitem{barak2013fixed}
O.~Barak, D.~Sussillo, R.~Romo, M.~Tsodyks, and L.~Abbott, ``From fixed points
  to chaos: three models of delayed discrimination,'' {\em Progress in
  neurobiology}, vol.~103, pp.~214--222, 2013.

\bibitem{carnevale2015dynamic}
F.~Carnevale, V.~de~Lafuente, R.~Romo, O.~Barak, and N.~Parga, ``Dynamic
  control of response criterion in premotor cortex during perceptual detection
  under temporal uncertainty,'' {\em Neuron}, vol.~86, no.~4, pp.~1067--1077,
  2015.

\bibitem{song2016training}
H.~F. Song, G.~R. Yang, and X.-J. Wang, ``Training excitatory-inhibitory
  recurrent neural networks for cognitive tasks: a simple and flexible
  framework,'' {\em PLoS computational biology}, vol.~12, no.~2, p.~e1004792,
  2016.

\bibitem{chaisangmongkon2017computing}
W.~Chaisangmongkon, S.~K. Swaminathan, D.~J. Freedman, and X.-J. Wang,
  ``Computing by robust transience: how the fronto-parietal network performs
  sequential, category-based decisions,'' {\em Neuron}, vol.~93, no.~6,
  pp.~1504--1517, 2017.

\bibitem{yang2019task}
G.~R. Yang, M.~R. Joglekar, H.~F. Song, W.~T. Newsome, and X.-J. Wang, ``Task
  representations in neural networks trained to perform many cognitive tasks,''
  {\em Nature neuroscience}, vol.~22, no.~2, pp.~297--306, 2019.

\bibitem{kim2021strong}
R.~Kim and T.~J. Sejnowski, ``Strong inhibitory signaling underlies stable
  temporal dynamics and working memory in spiking neural networks,'' {\em
  Nature Neuroscience}, vol.~24, no.~1, pp.~129--139, 2021.

\bibitem{serrano2022TID}
L.~Serrano-Fern{\'a}ndez, M.~Beir{\'a}n, and N.~Parga, ``Emergent perceptual
  biases from state-space geometry in spiking recurrent neural networks trained
  to discriminate time intervals,'' {\em bioRxiv}, 2022.

\bibitem{yamins2016using}
D.~L. Yamins and J.~J. DiCarlo, ``Using goal-driven deep learning models to
  understand sensory cortex,'' {\em Nature neuroscience}, vol.~19, no.~3,
  pp.~356--365, 2016.

\bibitem{dayan2005theoretical}
P.~Dayan and L.~F. Abbott, {\em Theoretical neuroscience: computational and
  mathematical modeling of neural systems}.
\newblock MIT press, 2005.

\bibitem{lecun2015deep}
Y.~LeCun, Y.~Bengio, and G.~Hinton, ``Deep learning,'' {\em nature}, vol.~521,
  no.~7553, pp.~436--444, 2015.

\bibitem{sompolinsky1988chaos}
H.~Sompolinsky, A.~Crisanti, and H.-J. Sommers, ``Chaos in random neural
  networks,'' {\em Physical review letters}, vol.~61, no.~3, p.~259, 1988.

\bibitem{lynn2020humans}
C.~W. Lynn and D.~S. Bassett, ``How humans learn and represent networks,'' {\em
  Proceedings of the National Academy of Sciences}, vol.~117, no.~47,
  pp.~29407--29415, 2020.

\bibitem{bishop2006pattern}
C.~M. Bishop and N.~M. Nasrabadi, {\em Pattern recognition and machine
  learning}.
\newblock Springer, 2006.

\bibitem{linsker1988self}
R.~Linsker, ``Self-organization in a perceptual network,'' {\em Computer},
  vol.~21, no.~3, pp.~105--117, 1988.

\bibitem{atick1990towards}
J.~J. Atick and A.~N. Redlich, ``Towards a theory of early visual processing,''
  {\em Neural computation}, vol.~2, no.~3, pp.~308--320, 1990.

\bibitem{nadal1994nonlinear}
J.-P. Nadal and N.~Parga, ``Nonlinear neurons in the low-noise limit: a
  factorial code maximizes information transfer,'' {\em Network: Computation in
  neural systems}, vol.~5, no.~4, pp.~565--581, 1994.

\bibitem{barlow1961coding}
H.~B. Barlow, ``The coding of sensory messages,'' {\em Current problems in
  animal behavior}, 1961.

\bibitem{nadal1994duality}
J.-P. Nadal and N.~Parga, ``Duality between learning machines: a bridge between
  supervised and unsupervised learning,'' {\em Neural Computation}, vol.~6,
  no.~3, pp.~491--508, 1994.

\bibitem{richards2019dendritic}
B.~A. Richards and T.~P. Lillicrap, ``Dendritic solutions to the credit
  assignment problem,'' {\em Current opinion in neurobiology}, vol.~54,
  pp.~28--36, 2019.

\bibitem{minsky1961steps}
M.~Minsky, ``Steps toward artificial intelligence,'' {\em Proceedings of the
  IRE}, vol.~49, no.~1, pp.~8--30, 1961.

\bibitem{hebb2005organization}
D.~O. Hebb, {\em The organization of behavior: A neuropsychological theory}.
\newblock Psychology Press, 2005.

\bibitem{roelfsema2018control}
P.~R. Roelfsema and A.~Holtmaat, ``Control of synaptic plasticity in deep
  cortical networks,'' {\em Nature Reviews Neuroscience}, vol.~19, no.~3,
  pp.~166--180, 2018.

\bibitem{magee2020synaptic}
J.~C. Magee and C.~Grienberger, ``Synaptic plasticity forms and functions,''
  {\em Annual review of neuroscience}, vol.~43, pp.~95--117, 2020.

\bibitem{rumelhart1986learning}
D.~E. Rumelhart, G.~E. Hinton, and R.~J. Williams, ``Learning internal
  representations by error propagation,'' tech. rep., California Univ San Diego
  La Jolla Inst for Cognitive Science, 1985.

\bibitem{werbos1988generalization}
P.~J. Werbos, ``Generalization of backpropagation with application to a
  recurrent gas market model,'' {\em Neural networks}, vol.~1, no.~4,
  pp.~339--356, 1988.

\bibitem{williams1992simple}
R.~J. Williams, ``Simple statistical gradient-following algorithms for
  connectionist reinforcement learning,'' {\em Machine learning}, vol.~8,
  no.~3, pp.~229--256, 1992.

\bibitem{lillicrap2019backpropagation}
T.~P. Lillicrap and A.~Santoro, ``Backpropagation through time and the brain,''
  {\em Current opinion in neurobiology}, vol.~55, pp.~82--89, 2019.

\bibitem{lillicrap2020backpropagation}
T.~P. Lillicrap, A.~Santoro, L.~Marris, C.~J. Akerman, and G.~Hinton,
  ``Backpropagation and the brain,'' {\em Nature Reviews Neuroscience},
  vol.~21, no.~6, pp.~335--346, 2020.

\bibitem{whittington2019theories}
J.~C. Whittington and R.~Bogacz, ``Theories of error back-propagation in the
  brain,'' {\em Trends in cognitive sciences}, vol.~23, no.~3, pp.~235--250,
  2019.

\bibitem{mato1992generalization}
G.~Mato and N.~Parga, ``Generalization properties of multilayered neural
  networks,'' {\em Journal of Physics A: Mathematical and General}, vol.~25,
  no.~19, p.~5047, 1992.

\bibitem{wu2017towards}
L.~Wu, Z.~Zhu, {\em et~al.}, ``Towards understanding generalization of deep
  learning: Perspective of loss landscapes,'' {\em arXiv preprint
  arXiv:1706.10239}, 2017.

\bibitem{maass2002real}
W.~Maass, T.~Natschl{\"a}ger, and H.~Markram, ``Real-time computing without
  stable states: A new framework for neural computation based on
  perturbations,'' {\em Neural computation}, vol.~14, no.~11, pp.~2531--2560,
  2002.

\bibitem{jaeger2002adaptive}
H.~Jaeger, ``Adaptive nonlinear system identification with echo state
  networks,'' {\em Advances in neural information processing systems}, vol.~15,
  2002.

\bibitem{jaeger2004harnessing}
H.~Jaeger and H.~Haas, ``Harnessing nonlinearity: Predicting chaotic systems
  and saving energy in wireless communication,'' {\em science}, vol.~304,
  no.~5667, pp.~78--80, 2004.

\bibitem{van1996chaos}
C.~Van~Vreeswijk and H.~Sompolinsky, ``Chaos in neuronal networks with balanced
  excitatory and inhibitory activity,'' {\em Science}, vol.~274, no.~5293,
  pp.~1724--1726, 1996.

\bibitem{molgedey1992suppressing}
L.~Molgedey, J.~Schuchhardt, and H.~G. Schuster, ``Suppressing chaos in neural
  networks by noise,'' {\em Physical review letters}, vol.~69, no.~26, p.~3717,
  1992.

\bibitem{bertschinger2004real}
N.~Bertschinger and T.~Natschl{\"a}ger, ``Real-time computation at the edge of
  chaos in recurrent neural networks,'' {\em Neural computation}, vol.~16,
  no.~7, pp.~1413--1436, 2004.

\bibitem{rajan2010stimulus}
K.~Rajan, L.~Abbott, and H.~Sompolinsky, ``Stimulus-dependent suppression of
  chaos in recurrent neural networks,'' {\em Physical review e}, vol.~82,
  no.~1, p.~011903, 2010.

\bibitem{dubreuil2022role}
A.~Dubreuil, A.~Valente, M.~Beiran, F.~Mastrogiuseppe, and S.~Ostojic, ``The
  role of population structure in computations through neural dynamics,'' {\em
  Nature Neuroscience}, pp.~1--12, 2022.

\bibitem{sutton2018reinforcement}
R.~S. Sutton and A.~G. Barto, {\em Reinforcement learning: An introduction}.
\newblock MIT press, 2018.

\bibitem{moore1990efficient}
A.~W. Moore, ``Efficient memory-based learning for robot control,'' 1990.

\bibitem{kaelbling1998planning}
L.~P. Kaelbling, M.~L. Littman, and A.~R. Cassandra, ``Planning and acting in
  partially observable stochastic domains,'' {\em Artificial intelligence},
  vol.~101, no.~1-2, pp.~99--134, 1998.

\bibitem{rao2010decision}
R.~P. Rao, ``Decision making under uncertainty: a neural model based on
  partially observable markov decision processes,'' {\em Frontiers in
  computational neuroscience}, vol.~4, p.~146, 2010.

\bibitem{sarno2017dopamine}
S.~Sarno, V.~de~Lafuente, R.~Romo, and N.~Parga, ``Dopamine reward prediction
  error signal codes the temporal evaluation of a perceptual decision report,''
  {\em Proceedings of the National Academy of Sciences}, vol.~114, no.~48,
  pp.~E10494--E10503, 2017.

\bibitem{sarno2022dopamine}
S.~Sarno, M.~Beir{\'a}n, J.~Falc{\'o}-Roget, G.~Diaz-deLeon, R.~Rossi-Pool,
  R.~Romo, and N.~Parga, ``Dopamine firing plays a dual role in coding reward
  prediction errors and signaling motivation in a working memory task,'' {\em
  Proceedings of the National Academy of Sciences}, vol.~119, no.~2,
  p.~e2113311119, 2022.

\bibitem{sutton1999policy}
R.~S. Sutton, D.~McAllester, S.~Singh, and Y.~Mansour, ``Policy gradient
  methods for reinforcement learning with function approximation,'' {\em
  Advances in neural information processing systems}, vol.~12, 1999.

\bibitem{fremaux2010functional}
N.~Fr{\'e}maux, H.~Sprekeler, and W.~Gerstner, ``Functional requirements for
  reward-modulated spike-timing-dependent plasticity,'' {\em Journal of
  Neuroscience}, vol.~30, no.~40, pp.~13326--13337, 2010.

\bibitem{schultz2000neuronal}
W.~Schultz and A.~Dickinson, ``Neuronal coding of prediction errors,'' {\em
  Annual review of neuroscience}, vol.~23, no.~1, pp.~473--500, 2000.

\bibitem{vasilaki2009spike}
E.~Vasilaki, N.~Fr{\'e}maux, R.~Urbanczik, W.~Senn, and W.~Gerstner,
  ``Spike-based reinforcement learning in continuous state and action space:
  when policy gradient methods fail,'' {\em PLoS computational biology},
  vol.~5, no.~12, p.~e1000586, 2009.

\bibitem{fiete2006gradient}
I.~R. Fiete and H.~S. Seung, ``Gradient learning in spiking neural networks by
  dynamic perturbation of conductances,'' {\em Phys. Rev. Lett.}, vol.~97,
  p.~048104, Jul 2006.

\bibitem{hoerzer2014emergence}
G.~M. Hoerzer, R.~Legenstein, and W.~Maass, ``{Emergence of Complex
  Computational Structures From Chaotic Neural Networks Through
  Reward-Modulated Hebbian Learning},'' {\em Cerebral Cortex}, vol.~24,
  pp.~677--690, 11 2012.

\bibitem{legenstein2010reward}
R.~Legenstein, S.~M. Chase, A.~B. Schwartz, and W.~Maass, ``A reward-modulated
  hebbian learning rule can explain experimentally observed network
  reorganization in a brain control task,'' {\em Journal of Neuroscience},
  vol.~30, no.~25, pp.~8400--8410, 2010.

\bibitem{wierstra2009recurrent}
D.~Wierstra, A.~Förster, J.~Peters, and J.~Schmidhuber, ``{Recurrent policy
  gradients},'' {\em Logic Journal of the IGPL}, vol.~18, pp.~620--634, 09
  2009.

\bibitem{chung2014empirical}
J.~Chung, C.~Gulcehre, K.~Cho, and Y.~Bengio, ``Empirical evaluation of gated
  recurrent neural networks on sequence modeling,'' {\em arXiv preprint
  arXiv:1412.3555}, 2014.

\bibitem{kiani2008bounded}
R.~Kiani, T.~D. Hanks, and M.~N. Shadlen, ``Bounded integration in parietal
  cortex underlies decisions even when viewing duration is dictated by the
  environment,'' {\em Journal of Neuroscience}, vol.~28, no.~12,
  pp.~3017--3029, 2008.

\bibitem{kim2018learning}
C.~M. Kim and C.~C. Chow, ``Learning recurrent dynamics in spiking networks,''
  {\em Elife}, vol.~7, p.~e37124, 2018.

\bibitem{nicola2017supervised}
W.~Nicola and C.~Clopath, ``Supervised learning in spiking neural networks with
  force training,'' {\em Nature communications}, vol.~8, no.~1, pp.~1--15,
  2017.

\bibitem{bellec2020solution}
G.~Bellec, F.~Scherr, A.~Subramoney, E.~Hajek, D.~Salaj, R.~Legenstein, and
  W.~Maass, ``A solution to the learning dilemma for recurrent networks of
  spiking neurons,'' {\em Nature communications}, vol.~11, no.~1, pp.~1--15,
  2020.

\bibitem{guerguiev2017towards}
J.~Guerguiev, T.~P. Lillicrap, and B.~A. Richards, ``Towards deep learning with
  segregated dendrites,'' {\em ELife}, vol.~6, p.~e22901, 2017.

\bibitem{sacramento2018dendritic}
J.~Sacramento, R.~Ponte~Costa, Y.~Bengio, and W.~Senn, ``Dendritic cortical
  microcircuits approximate the backpropagation algorithm,'' {\em Advances in
  neural information processing systems}, vol.~31, 2018.

\bibitem{payeur2021burst}
A.~Payeur, J.~Guerguiev, F.~Zenke, B.~A. Richards, and R.~Naud,
  ``Burst-dependent synaptic plasticity can coordinate learning in hierarchical
  circuits,'' {\em Nature neuroscience}, vol.~24, no.~7, pp.~1010--1019, 2021.

\bibitem{kording2001supervised}
K.~P. K{\"o}rding and P.~K{\"o}nig, ``Supervised and unsupervised learning with
  two sites of synaptic integration,'' {\em Journal of computational
  neuroscience}, vol.~11, no.~3, pp.~207--215, 2001.

\bibitem{Masse2022.05.09.491102}
N.~Y. Masse, M.~C. Rosen, D.~Y. Tsao, and D.~J. Freedman, ``Flexible cognition
  in context-modulated reservoir networks,'' {\em bioRxiv}, 2022.

\bibitem{parga1986ultrametric}
N.~Parga and M.~A. Virasoro, ``The ultrametric organization of memories in a
  neural network,'' {\em Journal de Physique}, vol.~47, no.~11, pp.~1857--1864,
  1986.

\bibitem{renart1999backward}
A.~Renart, N.~Parga, and E.~T. Rolls, ``Backward projections in the cerebral
  cortex: implications for memory storage,'' {\em Neural Computation}, vol.~11,
  no.~6, pp.~1349--1388, 1999.

\bibitem{renart2001model}
A.~Renart, R.~Moreno, J.~de~la Rocha, N.~Parga, and E.~T. Rolls, ``A model of
  the it-pf network in object working memory which includes balanced persistent
  activity and tuned inhibition,'' {\em Neurocomputing}, vol.~38,
  pp.~1525--1531, 2001.

\bibitem{brunel2001effects}
N.~Brunel and X.-J. Wang, ``Effects of neuromodulation in a cortical network
  model of object working memory dominated by recurrent inhibition,'' {\em
  Journal of computational neuroscience}, vol.~11, no.~1, pp.~63--85, 2001.

\bibitem{yang2021towards}
G.~R. Yang and M.~Molano-Maz{\'o}n, ``Towards the next generation of recurrent
  network models for cognitive neuroscience,'' {\em Current Opinion in
  Neurobiology}, vol.~70, pp.~182--192, 2021.

\end{thebibliography}

\end{document}